\documentclass[preprint, 3p, authoryear]{elsarticle}

\usepackage{amsmath}
\usepackage{amssymb}
\usepackage{graphicx}
\usepackage{hyperref}
\usepackage{setspace}
\usepackage{ulem}
\usepackage{tikz}
\usepackage{pgfplots}
\usepgfplotslibrary{fillbetween}
\usetikzlibrary{arrows.meta,automata,topaths,angles}
\usetikzlibrary{decorations.pathmorphing}
\usetikzlibrary{decorations.markings}  
\usetikzlibrary{patterns}
\usepackage{subcaption}
\usepackage{hhline}
\numberwithin{equation}{section}
\usepackage{lineno}
\newdefinition{rmk}{Remark}

\newcommand{\T}{\mathcal{T}}

\tikzset{decorated arrows/.style={
    postaction={
        decorate,
        decoration={
            markings,
            mark=at position 0.5 with {\arrow{stealth}}
            }
        },
    }
}

\journal{}
\begin{document}
\begin{frontmatter}

\title{Explicit Runge-Kutta algorithm to solve non-local equations with memory effects: case of the Maxey-Riley-Gatignol equation}

\author[]{Divya Jaganathan}
\author[]{Rama Govindarajan}
\author[]{Vishal Vasan}
\ead{vishal.vasan@icts.res.in}

\address{International Centre for Theoretical Sciences, Tata Institute of Fundamental Research, Shivakote, Bengaluru 560089, India}

\begin{abstract}
A standard approach to solve ordinary differential equations, when they describe dynamical systems, is to adopt a Runge-Kutta or related scheme. Such schemes, however, are not applicable to the large class of equations which do not constitute dynamical systems. In several physical systems, we encounter integro-differential equations with memory terms where the time derivative of a state variable at a given time depends on all past states of the system. Secondly, there are equations whose solutions do not have well-defined Taylor series expansion. The Maxey-Riley-Gatignol equation, which describes the dynamics of an inertial particle in  nonuniform and unsteady flow, displays both challenges. We use it as a test bed to address the questions we raise, but our method may be applied to all equations of this class. We show that the Maxey-Riley-Gatignol equation can be embedded into an extended Markovian system which is constructed by introducing a new dynamical co-evolving state variable that encodes memory of past states. We develop a Runge-Kutta algorithm for the resultant Markovian system. The form of the kernels involved in deriving the Runge-Kutta scheme necessitates the use of an expansion in powers of $t^{1/2}$. Our approach naturally inherits the benefits of standard time-integrators, namely a constant memory storage cost, a linear growth of operational effort with simulation time, and the ability to restart a simulation with the final state as the new initial condition. 
\end{abstract}
\end{frontmatter}

\section{Introduction}
\label{sec1}
Runge-Kutta methods and other iterative time-integrators of this class are extremely powerful techniques to solve ordinary differential equations (ODEs) that describe dynamical systems. However, several physical systems evolve non-locally in time and thus are not directly expressible as dynamical systems. Memory effects appear when the present state of a system depends explicitly on its past states, for instance, when the present state depends on the full trajectory of the solution up to the current time. Such memory effects pose a challenge for numerical simulations since standard algorithms are not designed for them. Another challenge, that often goes hand-in-hand with non-locality, but can also occur by itself in differential equations, is when the solution cannot be expressed in Taylor series, for instance, when it scales with fractional powers of the independent variable. In such cases, the solution and/or terms in the differential equation may not be differentiable, which violates a key hypothesis in the derivation of standard numerical schemes. The purpose of this paper is to address these challenges and to provide accurate Runge-Kutta algorithm for this class of problems. As a case in point we discuss the Maxey-Riley-Gatignol (MRG) equation, but our ideas and method can be extended to solve a large class of equations with a memory term. We characterise such equations as those that, when written in \textit{Duhamel} form, possess a kernel with a particular `spectral representation'. We will define the term spectral representation later in this introduction.

In the rest of the introduction, we describe the physical context of the MRG equation, the challenges posed by the memory effects, and the various approaches that have been taken so far to numerically solve the equation. The technical challenges faced in the numerical treatment are common to all equations with memory effects, and the MRG equation provides a concrete setting to address them. In addition, the MRG equation itself also poses other specific difficulties, which may not afflict all other equations with memory effects, but which might appear in other physical systems. We end the introduction with a high-level description of our proposed novel approach. The aim is to convince the reader that though the discussion may be motivated by the MRG equation, the approach is applicable more broadly.

The MRG equation describes the motion of an isolated, finitely-small particle in an unsteady and nonuniform flow, $u=u(x,t)$, in the Stokesian limit (where the Reynolds number of the particle as well as that of the shear across it are small). It is a common Lagrangian approach used to model dilute suspensions of spherical inertial particles in flows in a wide range of geophysical, environmental and industrial applications such as plankton in the ocean, cloud water droplets in air, and industrial sedimentation. The instantaneous state of a particle in such systems is given by its position and velocity, $[y(t),v(t)]^T$. Given its initial state $[y_0, v_0]^T$, the particle evolves according to the following set of non-dimensional equations: 
\begin{subequations}{\label{Eq:nfMRG}}
    \begin{align}
        \frac{dy}{dt} &= v(t), \label{Eq:nfMRG1}\\
        \frac{dv}{dt} +\alpha w + \gamma \Big( \frac{w_0}{\sqrt{\pi t}} + \int_0^t \frac{dw(\tau)/d\tau }{\sqrt{\pi(t-\tau)}} \:d\tau\Big) &=\frac{1}{R} \frac{D}{Dt}\Big(u(y(t),t)+\frac{a^2}{30L^2}\nabla^2 u(y(t),t)\Big) -\Big(\frac{1}{R}-1\Big) g, \label{Eq:nfMRG2}
    \end{align}
\end{subequations}
where 
\begin{equation}
    w(t) := v(t)-u(y(t),t) - \frac{a^2}{6L^2} \nabla^2 u(y(t),t)~.
\end{equation}
Eq. \ref{Eq:nfMRG2} is called the Maxey-Riley-Gatignol equation \citep{Maxey1983, gatignol1983,auton1988}, or often just the Maxey-Riley equation. It comprises a balance of forces for a small sphere in unsteady Stokes flow: (in their order of appearance after $dv/dt$) the quasi-steady Stokes drag, the Basset-Boussinesq memory force \citep{Basset1888, Boussinesq1885}, the added mass and pressure drag, and external forces (here, the gravitational force). The term on the left-hand side of Eq. \ref{Eq:nfMRG2} proportional to $\gamma$ is called the \emph{memory term} and represents non-local behaviour in $t$. The variables in Eq. \ref{Eq:nfMRG} correspond to physical variables which have been non-dimensionalised by a characteristic length scale $L$, a characteristic timescale $T$, and the velocity scale, $U=L/T$. The dimensionless parameters are
\begin{equation}{\label{Def:RSag}}
    R = \frac{2 \rho_p + \rho_f}{3\rho_f}, \quad S = \frac{a^2}{3\nu T}, \quad \alpha = \frac{1}{RS}, \quad\gamma = \sqrt{\frac{3}{R^2 S}}~,
\end{equation}
where $\rho_f$ and $\rho_p$ are the densities of the fluid and particle respectively, $\nu$ is the fluid kinematic viscosity and $g$ is the non-dimensional gravitational acceleration. The terms proportional to $\nabla^2 u$ are the Fax\'en correction terms which account for the non-uniformity in the flow across the particle's length. The time derivative $Du/Dt$ is the material derivative following the path of the fluid defined as $\partial u/\partial t + u\cdot \nabla u$, while the time derivative $du/dt$ (appearing in the memory term) is taken along the particle trajectory, and is defined as $du/dt = \partial u/\partial t + v(t) \cdot \nabla u$.\\
\indent The MRG equation belongs to the general class of fractional differential equations (FDEs). Indeed, the memory term is a fractional half-derivative in $t$, defined in the Riemann-Liouville sense
by 
\begin{equation}{\label{Def:RLder}}
     \frac{d^{1/2} w}{dt^{1/2}} := \frac{w_0}{\sqrt{\pi t}} + \int_0^t \frac{dw(\tau)/d\tau }{\sqrt{\pi(t-\tau)}} \:d\tau\:.
\end{equation}
The convolution operation in the above expression introduces non-locality in time. The slowly-decaying  $1/\sqrt t$-kernel of the convolution enforces a strong dependence on  long-past states of the particle. The weak singularity of the kernel acts in a particular way that the current state has the most `vivid' effect while the memory of earlier states slowly `fades away' as time elapses. Note that as a result of the memory term, the MRG equation no longer describes a dynamical system \citep{Farazmand15}. Indeed the state of the particle, namely the particle position and velocity at any given instant, is insufficient to uniquely determine its subsequent evolution in position-velocity space. 

For the rest of the paper, to simplify the explanation, we omit the Fax\'en corrections (they may be reintroduced as additional forcing terms under the same approach) and rewrite the MRG equation for the particle slip velocity, $w(t)=v(t)-u(y(t),t)$, in a compact form as:
\begin{subequations}{\label{Eq:compactMRG}}
\begin{align}
    \frac{dy}{dt} &= w + u(y(t),t)~, \label{Eq:compactMRG1}\\
    \frac{dw}{dt} &= - \alpha w - \gamma \frac{d^{1/2}w}{dt^{1/2}} + N(w(t), t) \label{Eq:compactMRG2},
\end{align}
\end{subequations}
where the memory term appears as a half-derivative in time, initial state is $y(0)=y_0, w(0)=w_0$, and we have bundled the nonlinear and external forces into a newly defined function, $N$:
\begin{equation*}
    N(w(t),t) = \Big(\frac{1}{R}-1\Big)\Big(\frac{Du}{Dt}(y(t),t) - g\Big) - w\cdot\nabla u(y(t),t)~.
\end{equation*}
Note that $u(x,t)$ and functions of it are evaluated at the present particle position, $y(t)$ which makes the system nonlinear. We assume the necessary smoothness properties for the underlying flow field $u(x,t)$. We remind the reader that while we address the MRG equation (Eq. \ref{Eq:compactMRG2}) and its numerical treatment in the rest of the paper, the equation for the particle position (Eq. \ref{Eq:compactMRG1}) is also co-evolved. Owing to its local nature, the particle position can be integrated, once its velocity is known, using any standard numerical scheme. Hence we do not discuss it any further. 

We are concerned with the numerical integration of the MRG equation. If we neglected the memory term (\textit{i.e.} set $\gamma=0$ in Eq. \ref{Eq:compactMRG}), the MRG equation  would reduce to a dynamical system described by an ODE which is readily solved numerically using standard time-integrators. It is due to this important simplification that the memory term is often neglected in particle dynamics computations, though such neglect may be unjustified \citep{haller2019}. To solve the MRG equation correctly, we must include the non-local memory term.  In other words, to accurately simulate the physics of an inertial particle, even an approximate numerical rule for the evolution of the particle state must depend on all prior states. Evidently, this precludes the use of standard time-integrators which are based on a local evolution rule. 

\indent When we solve an equation with non-locality in time using a numerical method prescribed on a time-grid $\{t_n = n\Delta t \}$, we anticipate the following computational costs: $(i)$ since at every time instant we need to perform an integral from the initial time to the present, the operational costs grow quadratically with simulation time instead of the more typical linear growth, and $(ii)$ since we have to store all past states, the memory storage requirement grows linearly with time instead of staying constant. This is also reflected in a related cost to restart the simulation from an arbitrary state: the state of the particle at a given time can no longer be prescribed as the new initial condition to restart the simulation, instead the entire trajectory up to the current time needs to be provided.  As a result of these costs, long-time and multi-particle simulations become prohibitively expensive. Moreover, the form of the memory term (\ref{Def:RLder}) warrants some care when handling the singular kernel numerically. 

Most existing numerical approaches to solve the MRG equation address the singularity issue by appealing to its integrability property \citep{brush1964, vanhinsberg2011, daitche2013}. To address the growing computational costs, a class of window-based approaches has been developed. This class of methods  splits the memory-term integral into two: one over the recent past and another over the distant past. Different methods  distinguish themselves in their treatment of the memory kernel in the two time-intervals. The $1/\sqrt{t}-$kernel is retained in the time-interval spanning the recent-past where an appropriate quadrature rule mindful of the singularity is used. Often, the kernel in the distant past is set to zero \citep{dorganloth2007, bombardelli2008}, effectively truncating the memory term. Alternatively, a fast-converging kernel using a combination of exponential functions is sought as an approximation to the kernel in the distant past \citep{vanhinsberg2011, parmar2018}. These approaches  control the operational costs in the following way: the quadratic growth persists for a predetermined length of the simulation time while a linear growth is realised after that. But, these approaches involve the introduction of new, and somewhat arbitrary, parameters, for example, the length of the `recent' time-interval and the number of exponentials in the approximation. These parameters need to be optimised, and that is often problem-specific. 

\subsection{An overview of the proposed method}\label{sec:overview}
\indent Our approach is a departure from those mentioned above. We pose the fundamentally non-local problem as an equivalent local-in-time problem. We then adapt standard time-integrators to solve the resultant local problem to reduce the various computational costs.

We embed the non-local (non-Markovian) MRG equation into a local (Markovian) system in an extended state-space. We show the construction of such an extended space by suitably introducing co-evolving state variable(s) in a natural manner. Similar embedding procedures are known in the literature in other contexts of classical and quantum non-Markovian processes \citep{markovianEmbedding_Siegle2011, markovianEmbedding1,markovianEmbedding2,markovianEmbedding_campbell}. A Markovian embedding was implicitly realised for the MRG equation  in \cite{sgp2019} where Eq. \ref{Eq:compactMRG2} was interpreted as an unsteady Robin boundary condition for the 1D diffusion PDE on a half-line. The authors exploited the fact that the Dirichlet-Neumann map to the diffusion problem is the Riemann-Liouville half-derivative (up to a sign). This allowed them to replace the non-local memory term in the Robin boundary condition (equivalently the MRG) with a local-in-time flux/Neumann term. The resulting numerical method essentially amounted to a \textit{fully-implicit} integration scheme wherein the Chebyshev coefficients of an approximate solution over the interval $[t_n,t_n + \Delta t]$ are numerically computed using a Newton iteration at every time-step.

The approach of \cite{sgp2019} employed a connection between the MRG equation and a PDE boundary-value problem (in this case the diffusion equation). In the present work, we dispose of this requirement and ultimately obtain  {\textit{explicit}} integration schemes of various orders of accuracy. We only require a particularly convenient `spectral representation' of the solution to the linear problem. Though this spectral representation is indeed a consequence of the connection to the diffusion equation, the essential requirement to eliminate memory effects is the existence of spectral representations and not necessarily a connection to some PDE boundary-value problem. 

A function $\chi(t)$ is said to have a {\textit{spectral representation}} if
\[\chi(t) = \int_{\Gamma} e^{f(k)t}g(k)dk\]
where $\Gamma$ is some contour over the $k-\mathbb{C}$ plane for some functions $f,g$ of the complex variable $k$. Typically, $f,g$ are analytic functions. An immediate consequence of this representation is our ability to embed the MRG equation into a dynamical system with a `larger' state description. To see this, we consider the MRG equation written as a nonlinear integral equation. The precise construction will be discussed later, but for now it suffices to note that differential equations can be written in integral form and this is often the starting point to develop numerical methods. Suppose $w:[0,\infty)\to\mathbb{R}$ is a sufficiently smooth solution to the integral equation
\[w(t) = \chi(t) w_0  + \int_0^{t} \chi(t-\tau)N(w(\tau),\tau)d\tau\]
where $N$ is a nonlinear function of $w$. If $\chi(t+\tau) = \chi(t)\chi(\tau)$ then one may ensure the integral need never be over an interval of length larger than $\Delta t$ (see Section \ref{sec2} for details). For equations that involve memory effects, it is often the case that $\chi(t+\tau)\neq \chi(t)\chi(\tau)$. In this latter case, $w(t)$ depends on $w(\tau)$ \textit{for all }$\tau\leq t$, \textit{i.e.}, the solution depends on the full trajectory and not just the present state. A direct discretisation of such integral equations was considered by \cite{Garrappa2011} for FDEs. In a later work, \cite{garrappa2012stability} provided Runge-Kutta type integrators for generic multi-term FDEs. However, these methods did not address the growing memory cost associated with the non-locality underpinning the FDE.

The spectral representation of $\chi(t)$ is one way to address the rising memory cost. Using the spectral representation for $\chi$ and up to a switch in the order of integration, we find the integral equation for $w(t)$ can be written as
\[w(t) = w_0\int_\Gamma e^{f(k)t}g(k)dk + \int_{\Gamma} \int_0^t e^{f(k)(t-\tau)}g(k)N(w(\tau),\tau)d\tau\:dk~. \]
If we define 
\[H(k,t) = w_0e^{f(k)t}g(k) + \int_0^t e^{f(k)(t-\tau)}g(k)N(w(\tau),\tau)d\tau~, \]
then we infer 
\[\frac{dH}{dt} - f(k)H = g(k)N(w(t),t),\quad H(k,0) = w_0g(k) \quad\mbox{ with } w(t) = \int_{\Gamma}H(k,t)dk~,\] 
which is a local-in-time reformulation of the integral equation for $w$ in terms of a new state-variable $H(\cdot,t)$ (considered as a function of $k\in\Gamma\subset \mathbb{C}$). Note the embedding into a dynamical system naturally arises from the spectral representation of the kernel of the integral equation $\chi(t)$. 

As mentioned earlier, Markovian embeddings for non-local equations have been considered elsewhere in different contexts. For the MRG equation, a Markovian embedding was implicit in the work of \cite{sgp2019} but it was neither recognised nor utilised to construct numerical schemes.  Attempts to express the memory term in Eq. \ref{Eq:nfMRG2} as a combination of exponentials to aid in the development of numerical schemes have been made before \citep{vanhinsberg2011,parmar2018}. In such works, the memory term is approximated by a combination of exponentials, whereas the strategy explained above is based on an \emph{exact} representation. Furthermore, in our approach, we employ a spectral representation for the solution to the linear problem and not the memory term in Eq. \ref{Eq:nfMRG2}. Essentially, this implies that we approximate a function that behaves like $\sqrt{t}$ for small $t$, as opposed to the aforementioned works which approximate functions that behave like $1/\sqrt{t}$.

\indent The resulting equation for $H$,
\begin{equation}{\label{eq:Heqn}}
    \frac{dH}{dt} - f(k)H = g(k)N(w(t),t),\:k\in\Gamma, \quad \mbox{where} \quad w(t) =  \int_{\Gamma}H(k,t)dk
\end{equation}
is indeed a dynamical system but the state-space is infinite-dimensional. Typically, $\Gamma$ is a continuum and hence to derive a numerical method for the dynamical system we need to (a) expand $H(\cdot,t)$ in a basis, (b) truncate this expansion to obtain a finite-dimensional approximation. Once a basis of functions is chosen, the above equation is amenable to the exponential time differencing scheme of \citep{CM2002}. Recall the equation for $H$ was derived from an integral-equation version of the MRG equation. Moreover, the $H-$equation is \textit{local-in-time}. Thus, we can eliminate the rising memory costs if we can find a small number of basis functions that provide a sufficiently accurate approximate solution to the $H-$equation for all time. The number of terms in a basis-expansion for $H(\cdot,t)$ depends on how smooth $H$ is (as a function of $k$) and the integrability of $H(\cdot,t)$ over $\Gamma$. For the MRG equation, we find that a direct discretisation of the $H-$equation requires a large number of basis functions; too large to be of practical use (see Section \ref{sec6p3}). 

The numerical method proposed in this manuscript corresponds to a synthesis of the ideas presented in the above paragraphs. We employ a Markovian embedding but also employ generalisations of the exponential time differencing schemes. Our work can also be considered as a kind of window-based method that is motivated by the spectral representation. Our work closely follows the procedure of \cite{HO2005} to construct a Runge-Kutta algorithm for the resultant local system. Unlike the standard Runge-Kutta derivations which assume a Taylor series expansion for the unknown solution, we are forced to employ expansions in non-negative powers of $t^{1/2}$ for the solution. This is a consequence of the smoothness properties of the solution to the linear MRG equation. However, it also captures a feature of the physics, namely the role of the initial slip velocity $w_0$ in Eq. \ref{Def:RLder}.

We summarize the contributions of the work:
\begin{itemize}
        \item We propose a method to numerically integrate the non-Markovian MRG equation by embedding the equation into a larger Markovian system and deriving Runge-Kutta schemes for the resulting system of equations. 
        \item The method we propose results in a numerical scheme which
        \begin{enumerate}
            \item[(a)] is explicit and local in time, reflecting the causal nature of the underlying physical system,
            \item[(b)] has constant memory storage and simulation restart cost,
            \item[(c)] can handle non-zero initial condition without suffering loss of accuracy, and
            \item[(d)] can be derived with tunable order of accuracy in the numerical step-size $\Delta t$.
        \end{enumerate}
        \item We adapt the ideas of \cite{CM2002} to equations which do not possess a semigroup structure for the linear problem, while still maintaining a constant memory cost.
        \item Our method follows naturally from the existence of a spectral representation for the linear solution operator. Both the Markovian embedding and the need for expansions in powers of  $t^{1/2}$ in the Runge-Kutta derivation are consequences of the spectral representation relevant to the MRG equation.
        \item Our method generalises to other equations with memory effects subject to the existence of the relevant spectral representation (Section \ref{sec7}). We emphasize that the spectral representation for the linear solution (and its associated properties) is what dictates the development of the numerical method.
\end{itemize}
In the next few sections, we develop the ideas for our numerical procedure. The reader who simply wishes to know the specifics of the numerical method may skip to Section \ref{sec4} directly. In Section \ref{sec5} we present an error analysis for the Runge-Kutta schemes and in Section \ref{sec6} we present some numerical experiments to verify the expected convergence rates.

\section{The semigroup property and numerical integration}{\label{sec2}}
In the context of this manuscript, a real-valued function $S:\mathbb{R}^+\to\mathbb{R}$ is said to have the semigroup property if $S(t+\tau)=S(t)S(\tau),\:\forall t,\tau\in \mathbb{R}^+$ where the product on the right-hand side is the product of real numbers. The definition can be extended to matrix-valued functions of a real-parameter $t$ by interpreting the product as the matrix product. Additionally, one often sets $S(0)=1$ (the multiplicative identity) and requires some smoothness of the function $S$. We refer the reader to \cite{curtain2012introduction} for more rigorous definitions of the semigroup property for solutions to differential equations.

In this section, we highlight the importance of the semigroup property in deriving numerical integration schemes. The semigroup property is a property of the solution to the underlying equation which also plays a key role in maintaining constant memory costs while designing numerical methods. To this end, we first discuss the exponential time differencing method \citep{CM2002} where the semigroup property of the solution to the linear part of the ODE plays an important role. Consider Eq. \ref{Eq:compactMRG2} when $\gamma=0$. This leads to a first-order ODE for a scalar quantity $w$ that evolves according to
\begin{equation}\label{slODE}
\frac{dw}{dt} = -\alpha w + N(w,t), \text{ }w(0) = w_0,
\end{equation}
where $N$ is a nonlinear function of the solution and the parameter $\alpha$ is a real quantity. The solution to the linear part of the equation is $w_0\exp(-\alpha t)$. Hence we can identify the function $S(t)=\exp(-\alpha t)$ which evidently has the semigroup property. Using the Duhamel solution for the linear forced problem, the solution to \eqref{slODE} satisfies the following integral equation,
\begin{equation}\label{slODE:duhamel}
w(t) = e^{-\alpha t}w_0 + \int_0^t e^{-\alpha (t-\tau)} N(w(\tau),\tau) \: d\tau.
\end{equation} 
Since the solution to the linear problem involves exponential functions, the numerical methods derived based on Eq. \ref{slODE:duhamel} are called exponential time differencing (ETD) methods. 

A typical ETD method is constructed as follows. Let $w(t_n)$ denote the solution at the discrete time $t_n$ and $\Delta t = t_{n+1} - t_n$ be the step-size.  Since the exponential function has the semigroup property, \textit{i.e.}, $\exp(-\alpha \Delta t)\exp(-\alpha t_n)=\exp(-\alpha (t_n+\Delta t))$ for every $n$, we have
\begin{align}\label{slODE:discrete}
w(t_{n+1}) &= e^{-\alpha t_{n+1}} \:w_0 + \int_0^{t_{n+1}} e^{-\alpha (t_{n+1}-s)} N(w(\tau),\tau)\:  d\tau,
\end{align}
\begin{align*}
\implies w(t_{n+1}) &= e^{-\alpha \Delta t}\left[e^{-\alpha t_{n}} w_0 + \int_0^{t_n} e^{-\alpha (t_n-\tau)} N(w(\tau),\tau) \: d\tau\right] + \int_{t_n}^{t_{n+1}} e^{-\alpha (t_{n+1}-\tau)} N(w(\tau),\tau)\:  d\tau,
\end{align*}
\begin{align}\label{slODE:iterative}
\implies w(t_{n+1}) &= e^{-\alpha \Delta t}\: w(t_n) + \int_0^{\Delta t} e^{-\alpha (\Delta t-\tau)} N(w(t_n + \tau),t_n+\tau)\: d\tau~.
\end{align}
Note that the semigroup property has been exploited to write the local iterative rule Eq. \ref{slODE:iterative}. More precisely, we express the solution at $t_{n+1}$ in terms of the solution at the prior time $t_n$ and integral over its small $\Delta t$-neighbourhood, for all $n$. Since the integral is over a small interval, we can find numerical approximation by assuming sufficient smoothness of the functions $w$ and $N$. Note also that it is precisely the semigroup property of the linear solution that permits us to replace the integral over the larger interval $[0,t_{n+1}]$ by one over a smaller interval $[0,\Delta t]$.

Different approximations for the integral term in Eq. \eqref{slODE:iterative} containing the nonlinearity lead to different numerical schemes. If we denote the numerical approximation to the solution at time $t_n$ by $w_n$, a canonical explicit, multi-step scheme (such as the ETD-Adam Bashforth method) for a fixed choice of integer $m\leq n$ is given as follows,
\begin{equation}	
w_{n+1} = e^{-\alpha \Delta t} w_{n} + \Delta t \sum_{j=1}^m W_j N(w_{n+1-j},t_{n+1-j})
\nonumber.
\end{equation}
Here the choice of nodes and corresponding weights $(\{t_{n+1-j}\},\{W_{j}\})$ defines a scheme with a certain $p$-order local accuracy $|w_n - w(t_n)| \sim \mathcal{O}(\Delta t)^{p+1}$. On the other hand, a canonical explicit, multi-stage (single-step) scheme of $s$-stages such as the ETD-Runge Kutta method (ETD-RK) is given by,
\begin{equation*}
\begin{aligned}	
w_{nj} &= e^{-\alpha c_j \Delta t} w_{n} + \Delta t \sum_{l=1}^{j-1} \tilde{W}_{jl} N(w_{nl},\tau_{nl}),\\
w_{n+1} &= e^{-\alpha \Delta t} w_{n} + \Delta t \sum_{j=1}^s W_j N(w_{nj},\tau_{nj})~,\\
\end{aligned}
\end{equation*}
where $\{w_{nj}\}$ are the intermediate stages of the solution progressively constructed at times $\{ \tau_{nj} = t_n + c_j \Delta t\}$ for $0\leq c_j \leq 1$. The final leap to the solution at $t_{n+1}$ is constructed from these intermediate evaluations for different (non-unique) choices of weights $\{W_j, \tilde{W}_{jl}\}$. See \cite{CM2002} for expressions for the weights in either of the schemes mentioned above. Note that in either construction (ETD-RK or ETD-Adams Bashforth) there is a notion of locality in the information required to advance the solution; an $m\Delta t -$neighbourhood in the multi-step schemes and a $\Delta t-$neighbourhood in the multi-stage schemes - which is a fixed size in both cases. We say the multi-stage method is a numerical Markovian method since the information in $\Delta t$-neighbourhood required to compute $w_{n+1}$ is effectively constructed from only one prior information at $t_{n}$. In contrast to the multi-step methods where the first few iterations suffer from dearth of information leading to loss of accuracy (unless compensated for, say, with adaptive time-integration), a multi-stage method affords a uniform rule at all iterations. As our goal is to derive Markovian numerical schemes for \eqref{Eq:compactMRG} we prefer to derive explicit, multi-stage numerical schemes. 

\subsection{A generic system without the semigroup property}{\label{sec2p1}}
\indent Consider now an integral equation such as \eqref{slODE:duhamel} but where the linear solution \textit{does not} have the semigroup property. In other words, in the place of the exponential function, we have a function $\T(t)$ such that $\T(t+\tau)\neq \T(t)\T(\tau)$. Following a similar procedure as before, we arrive at the following relationship for every $n$,
\begin{equation}\label{slFDE:notiter}
\begin{aligned}
w(t_{n+1}) &= \T(t_{n+1}) w_0 + \int_0^{t_{n+1}} \T(t_{n+1}-\tau) N(w(\tau),\tau) \:d\tau,\\
&=\Big(\T(t_{n+1})w_0 + \int_0^{t_n} \T(t_{n+1}-\tau) N(w(\tau),\tau) \:d\tau\Big) + \int_0^{\Delta t} \T(\Delta t-\tau) N(w(t_n+\tau),t_n+\tau)\: d\tau,\\
&= \mathcal{R}_{\text{history}}(t_n;t_{n+1}) + \int_0^{\Delta t} \T(\Delta t-\tau) N(w(t_n+\tau),t_n+\tau)\: d\tau~.
\end{aligned}
\end{equation}
Subject to similar smoothness requirements as before, the integral-term over the interval $[0,\Delta t]$ can be replaced by numerical approximations. The  residual expression we denote by $\mathcal R_{\text{history}}(t_n;t_{n+1})$ represents the influence of the solution \textit{up to} time $t_n$ on the solution at time $t_{n+1}$. When $\T$ has the semigroup property, $\mathcal R_{\text{history}}(t_n;t_{n+1})=\T(\Delta t)w_n$.

In the absence of the semigroup property for $\T$, the term $\mathcal R_{\text{history}}$ does not simplify and contains an integral over the entire history of the solution. Note that the size of this integral is itself time-dependent. This non-local nature of $\mathcal{R}_{\text{history}}$ presents significant difficulty. To evolve the solution by just one additional time-step, we need to integrate the full trajectory up to the current time. In other words, we have no notion of locality, contrary to the previous case where $\exp(\circ)$ possessed the semigroup property. Comparing with Eq. (\ref{slODE:iterative}), the reader will realise that Eq. (\ref{slFDE:notiter}) lacks a local-in-time iterative rule to establish a numerical scheme. From a computational standpoint, an iterative rule that is the same for every time-step, ensures that the cost to perform each iterate is time-independent, i.e., remains the same at each time-step. This in principle is lost when $\T$ lacks the semigroup property.\\

{\rmk{
The loss of the semigroup property is not as uncommon as one might assume. Indeed consider the second-order ODE 
\begin{equation}
    \frac{d^2 w}{dt^2}-\alpha w = N(w(t),t).
\end{equation}
The solution can be written in integral form as
\[w(t) = C_1 e^{\sqrt\alpha t} + C_2 e^{-\sqrt\alpha t} + \frac{1}{2\sqrt \alpha} \int_0^t (e^{\sqrt\alpha(t-\tau)}-e^{-\sqrt\alpha(t-\tau)})N(w(\tau),\tau)\:d\tau ~.\]
In the above description,  $w$ is the sole state variable and we have a non-Markovian process. Equivalently, the solution for the linear equation lacks the semigroup property. The linear solution for the above Cauchy problem is a linear combination of exponential functions. Whereas a single exponential function satisfies the semigroup property, a \textit{linear combination} does not. A natural (and indeed most common approach) taken to study this example, analytically or numerically, is to write it as a Markovian system of two state-variables, $[w, dw/dt]^T$, each with its own dynamical evolution equation. We present this example to highlight the fact that functions lacking the semigroup property can arise naturally. But we also stress that a Markovian embedding is a natural and common approach to address these kinds of problems. Indeed our approach can also be thought of as a Markovian embedding but into a state-space of much higher dimension. 
}}

\subsection{The linear Maxey-Riley-Gatignol equation}{\label{sec2p2}}
Linear fractional differential equations (FDEs) possess solutions that lack the semigroup property. Indeed such solutions may belong to the class of functions known as Mittag-Leffler functions \citep{Garrappa2011} which are continuous and bounded for $t\geq 0$, however, they may not be differentiable at $t=0$. The MRG equation naturally falls under the category of FDEs. In this subsection, we show that the MRG equation serves as a model problem to illustrate the main ideas of the present work. In the remainder of the paper we provide specifics for the scalar MRG equation, but the considerations are not limited to the specific equation or its scalar form. 

\indent We recall the MRG equation written in compact form \ref{Eq:compactMRG},
\begin{equation*}
\frac{dw}{dt} + \gamma \frac{d^{1/2}}{dt^{1/2}}w + \alpha w = N(w,t),\quad  w(0) = w_0,
\end{equation*}
where $\alpha, \gamma$ are positive real numbers (see \ref{Def:RSag} for definition) and $w_0$ is the initial slip velocity. We begin by identifying the solution to the linear MRG equation. We denote by $\chi(t;\alpha,\gamma)$ the function satisfying the linear MRG equation with initial condition as $1$. Hence 
\[\frac{d\chi}{dt} + \gamma \frac{d^{1/2}}{dt^{1/2}}\chi + \alpha \chi = 0,\quad  \chi(0;\alpha,\gamma) = 1.\]
One may use the Laplace transform to obtain a representation of the function $\chi$
\begin{equation}\label{chiform: laplace}
\chi(t;\alpha,\gamma) = \mathcal{L}^{-1}\Big[\frac{1}{s+\gamma \sqrt{s} + \alpha} \Big](t)~ = \frac{i}{\pi}\int_{-\infty}^\infty \frac{e^{-k^2t} k}{-k^2+ik\gamma + \alpha}dk
\end{equation} 
 where the second equality (integral over $k$) is obtained by a suitable re-parameterization of the associated Bromwich contour for the inverse Laplace transform (see \ref{A1}). Depending on the sign of the discriminant of the quadratic expression in the denominator, $(\gamma^2 - 4\alpha)$, different closed-form expressions are obtained \citep{Langlois15}. We list some notable properties of $\chi(t;\alpha,\gamma)$:
\begin{itemize}
\item \textit{Property 1: Limiting forms and short-time behaviour}
\[\chi(t;\alpha,0) = e^{-\alpha t},\quad \chi(t;0,\gamma) =e^{\gamma^2 t} \text{erfc}(\gamma \sqrt{t})~.\]
When $\gamma=0$ we obtain the reduced MRG equation which corresponds to a dynamical system. Indeed note the semigroup property for $\chi(t;\alpha,0)$. When $\alpha=0$ we obtain a function belonging to the class of \textit{Mittag-Leffler functions}. This function behaves as a stretched exponential, of the form $\exp(ae^{t^b})$, for small $t$. In particular, we have the series expansion in powers of $t^{1/2}$ around $t=0$,
\begin{equation*}
    \chi(t \rightarrow 0;\alpha=0, \gamma) \sim 1 -  \frac{2}{\sqrt{\pi}}\gamma \sqrt{t} + \gamma^2  t + ...
\end{equation*} 
 Note even for $\alpha\neq 0$, the integral expression \eqref{chiform: laplace} indicates $d\chi/dt$ is unbounded as $t\to 0$ and hence $\chi(t)$ behaves singularly as $t$ approaches zero. This implies $\chi$ lacks smoothness at $t=0$. However, for $t>0$, it can be shown that a regular expansion in integer powers of $t$ exists. We will recall this property for the short-time expansion of the solution to the MRG equation in constructing our numerical scheme to accurately capture the singular behaviour. 
  
\item \textit{Property 2: Not a semigroup}
\[\chi(t_1;\alpha, \gamma) \chi(t_2;\alpha, \gamma) \neq \chi(t_1+t_2; \alpha, \gamma) \]
Since the semigroup property ought to hold for all $t\geq 0$, the above may be readily verified by evaluating $\chi$ at suitable $t_1,t_2$. This is true when $\alpha=0$ too. Thus, the MRG equation by itself cannot be staged for iterative numerical integration methods such as the Runge-Kutta methods. However, note that, in contrast, the reduced MRG (in the $\gamma=0$ limit) regains the semigroup property allowing the use of standard iterative numerical integration methods.
\end{itemize}

\indent Having identified the solution to the linear MRG equation, the solution to the MRG equation with nonlinear forcing can be expressed in the following integral equation form
\begin{equation}\label{MR:duhamel}
    w(t) = \chi(t;\alpha,\gamma)w_0 + \int_0^t \chi(t-\tau;\alpha,\gamma)N(w(\tau),\tau)\:d\tau ~.
\end{equation} 
Following the discussion in Section \ref{sec2p1}, due to the lack of semigroup property, we have for the solution $w(t)$ at time $t_{n+1}$ the following
\begin{subequations}\label{MRnoniter}
\begin{align}
    w(t_{n+1}) &= \mathcal{R}_{\text{history}}^{\chi}(t_n;t_{n+1})+\int_{0}^{\Delta t}\chi(\Delta t-\tau;\alpha, \gamma)N(w(t_n + \tau),t_n+\tau)\: d\tau, \\
\text{where,} \quad \mathcal{R}_{\text{history}}^{\chi}(t_n;t_{n+1}) &=\chi(t_{n+1};\alpha,\gamma)w_0 + \int_0^{t_n} \chi(t_{n+1}-\tau;\alpha,\gamma)N(w(\tau),\tau)\:d\tau~,\label{Rdef}
\end{align}
\end{subequations}
where we have fallen short of writing a local iterative rule similar to Eq. \ref{slODE:iterative}. Note that the above are exact statements and do not involve any numerical approximation yet.

\section{Markovian embedding in an inflated state}{\label{sec3}}
The expression for the solution at $t_{n+1}$ in \eqref{MRnoniter} contains a residual term (\ref{Rdef}) that captures the influence of the history of past states on $w(t_{n+1})$. In this Section, we show that we can introduce a co-evolving state variable, denoted by $H$ and with its own dynamical equation, that can be used to compute $\mathcal{R}_{\text{history}}^\chi$. The key point is the spectral representation for $\chi$, namely
\begin{equation}\label{chispectral}
\chi(t; \alpha, \gamma) = \frac{i}{\pi} \int_{-\infty}^{\infty}\frac{e^{-k^2 t} k}{-k^2 +ik\gamma + \alpha} \: dk,
\end{equation}
exists, and it allows for a Markovian embedding of the MRG equation.

Let $H(k,t)$ be a function of a real number $k$ and time $t$ related to the residual term  $\mathcal{R}_{\text{history}}^{\chi}$ in Eq. (\ref{Rdef}) in the following manner
\[\chi(t_{n+1}; \alpha, \gamma) w_0 + \int_0^{t_n} \chi(t_{n+1}-\tau; \alpha, \gamma) N(w(\tau),\tau) \: d\tau = \mathcal{R}_{\text{history}}^{\chi}(t_n;t_{n+1}) = \int_{-\infty}^{\infty} H(k,t_n) e^{-k^2 \Delta t} \: dk ~.\]
Substitution of the spectral representation of $\chi$, Eq. (\ref{chispectral}), into the definition of $\mathcal{R}^{\chi}_{\text{history}}$ yields the following definition for the newly introduced variable $H(k,t)$, 
\begin{equation}\label{Hdef}
H(k,t) := \frac{i}{\pi}\Big( \frac{k e^{-k^2t}}{-k^2+ik\gamma+\alpha} w_0 + \int_0^t \frac{ke^{-k^2(t-\tau)}}{-k^2+ik\gamma+\alpha}N(w(\tau),\tau) \: d\tau\Big).
\end{equation}
Since $H$ encodes the influence of past states on the current state $w(t_{n+1})$ we refer to it as the history function. $H$ is a complex-valued function of the real-number $k$ and so is, strictly speaking, an infinite-dimensional object. 

{\rmk{\label{twoHs}The careful reader will notice a subtle difference between the function $H$ introduced here and the one referred to in Section \ref{sec:overview}. Here, $H$ is introduced specifically to account for the influence of past states on the present state, whereas in Section \ref{sec:overview} the function $H$ accounts for all the states up to the present state through $w(t) = \int H(k,t)dk$. The upshot is that the history function $H$ introduced in this section is multiplied by $\exp(-k^2\Delta t)$ to determine $\mathcal R_{\text{history}}^\chi$. The rapid decay afforded by the Gaussian multiplier helps us achieve a \textit{smaller} finite-dimensional approximation for $H$. This keeps the overall memory cost lower than the case presented in Section \ref{sec:overview}. See also Section \ref{sec6p3}.}}

\indent From the form of $H$ in \eqref{Hdef} one can infer it has Markovian dynamics evolving according to the following ODE and initial condition,
\begin{equation}{\label{dyneq4H}}
\frac{ dH(k,t)}{dt} + k^2 H(k,t) = \frac{i}{\pi} \frac{k}{(-k^2+ik\gamma+\alpha)}N(w(t),t), \quad H(k,0) = \frac{i}{\pi} \frac{k}{(-k^2+ik\gamma+\alpha)} w_0~.
\end{equation}
Indeed the solution to the above initial-value problem, for each $k$, is  precisely the left-hand side of \eqref{Hdef}. Thus, upon the introduction of the history function as a new state variable which co-evolves with $w$, we obtain the complete (larger dimensional) state that evolves `locally' from $t_n$ to $t_{n+1}$ according to
\begin{subequations}\label{newstatedynamics}
\begin{align}
w(t_{n+1}) &= \int_{-\infty}^{\infty} H(k,t_n) e^{-k^2 \Delta t} \: dk + \int_0^{\Delta t} \chi(\Delta t-\tau; \alpha, \gamma) N(w(t_n+\tau),t_n+\tau) \:d\tau, \label{newstatedynamics1}\\
H(k,t_{n+1}) &= e^{-k^2 \Delta t} H(k,t_n) + \frac{i}{\pi} \frac{k}{(-k^2+ik\gamma+\alpha)} \int_0^{\Delta t} e^{-k^2(\Delta t-\tau)} N(w(t_n+\tau),t_n+\tau) \: d\tau \label{newstatedynamics2}~.
\end{align}
\end{subequations}
Once again we emphasize that there has been no numerical approximation up to this point. Equation \eqref{newstatedynamics} is an entirely equivalent way to write Eq. \eqref{MRnoniter} (which is simply the integral version of Eq. \ref{Eq:compactMRG}).

\indent We prefer to work with real-valued functions and note that $H(-k,t)=\overline{H(k,t)}$ (over-bar indicates complex conjugate, see Eq. (\ref{Hdef})). To this end, we introduce the following simplifying re-definitions of variables and functions 
\[ k\sqrt{\Delta t} \rightarrow k, \quad \text{Re}\Big(\frac{2}{\sqrt{\Delta t}} H(k/\sqrt{\Delta t}, t)\Big) \rightarrow H(k,t).\]
Next, we prefer working with the function $\chi(t;\alpha=0,\gamma)$ which has an explicit form in terms of the complimentary error function. This allows us to better handle integrals involving the history function that will appear in the subsequent sections. This motivates us to absorb the term proportional to $\alpha$ in Eq. \eqref{Eq:compactMRG} (the Stokes' drag) into the nonlinear term. Hence we also introduce the following definitions
\begin{align} 
\tilde{\gamma} = \gamma \sqrt{\Delta t} , \quad \chi(t;\gamma) := \chi(t; 0,\gamma) = e^{\gamma^2 t}\text{erfc}(\gamma \sqrt{t})~, \quad N_{\alpha}(w) = N(w) - \alpha w~.\label{gammaN:redef}
\end{align}

Under the above transformations, we rewrite the evolution rule for the new inflated state Eq. (\ref{newstatedynamics}) in the integral-equation form:
\begin{subequations}\label{newstatetransformed}
\begin{align}
w(t_{n+1}) &= \int_{0}^{\infty} H(k,t_n) e^{-k^2} \: dk + \int_0^{\Delta t} \chi(\Delta t-\tau; \tilde{\gamma}) N_{\alpha}(w(t_n+\tau),t_n+\tau) \:d\tau, \label{newstatetransformed1}\\
H(k,t_{n+1}) &= e^{-k^2 } H(k,t_n) + \frac{2}{\pi} \frac{\tilde{\gamma}}{(k^2+\tilde{\gamma}^2)} \int_0^{\Delta t} e^{-k^2(1-\tau/\Delta t)} N_{\alpha}(w(t_n+\tau),t_n+\tau) \: d\tau \label{newstatetransformed2}~.
\end{align}
\end{subequations}
where $H$ is now a real-valued function. One may also note on comparing Eq. (\ref{newstatetransformed1}) with Eq. (\ref{MRnoniter}) in the original framework that we have effectively replaced the non-locality in time with a local-in-time term. This of course has come at the cost of introducing Eq. (\ref{newstatetransformed2}). Thus, while the new (extended) state description $[w(t), H(k,t)]^T$ is larger compared to the original MRG equation, $w(t)$, the associated memory and computational costs to advance the state do not grow with simulation time. As evident from the above discussion, to obtain the new Markovian state (\textit{i.e.} find the Markovian embedding) we only need the spectral representation for $\chi$. In the next section, we derive numerical time-integration schemes to evolve the new Markovian state. 

\begin{figure}[!ht]
    \centering
    \subfloat[\centering \label{schematic1} ]{{\includegraphics[width=0.7\textwidth]{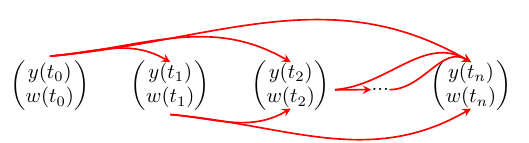} }}%
    \\
    \subfloat[\centering \label{schematic2} ]{{\includegraphics[width=0.7\textwidth]{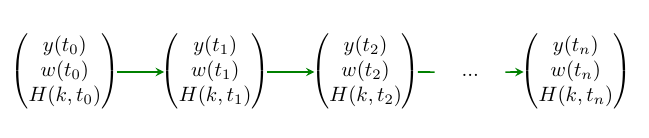} }}%
    \caption{A schematic for the Markovian embedding procedure for the MRG equation (Section \ref{sec3}). (a) Non-local inter-state interaction in the original representation Eq. \ref{Eq:compactMRG}. Interaction between no two states is the same. (b) Local interaction via Eq. \ref{newstatetransformed} due to inflated state description with the introduction of new 'history function', $H(k,t)$. All local interactions are identical. }
\end{figure}

\section{Explicit multi-stage time-integrator for the MRG equation}{\label{sec4}}

Let us summarise our progress so far. We begin with Eq. \eqref{Eq:compactMRG2} that prescribes the evolution of the slip velocity for a rigid spherical particle in a viscous fluid. We intend to find numerical approximations to this equation on a discrete grid in time with $\Delta t$ representing the temporal discretization. We recall the following notation,
\[\tilde{\gamma} = \gamma \sqrt{\Delta t} , \quad \chi(t;\gamma) = e^{\gamma^2 t}\text{erfc}(\gamma \sqrt{t})~, \quad N_{\alpha}(w) = N(w) - \alpha w~.\]
Note $w,N(w),N_\alpha(w)$ are typically $3-$dimensional real-valued vectors for the MRG equation. Based on the discussions in the previous sections, the solution to the MRG equation at time $t_{n+1}$ satisfies Eq. \eqref{newstatetransformed} where $H(k,t_n)$ is also a $3-$dimensional vector for each $k\in\mathbb{R}$. A schematic summarising the non-local interaction, and its representation as a dynamical system as a result of the embedding procedure is given in Figures \ref{schematic1} and \ref{schematic2} respectively.

\indent In this section, we outline the multi-stage time-integrator for the MRG equation rewritten under the Markovian embedding according to Eq. (\ref{newstatetransformed}), and provide details of the derivation in Section \ref{sec5}. We introduce grid functions $w_{n}, H_n(k)$ which are the numerical approximations of their exact counterparts $w(t_n), H(k,t_n)$ respectively. In addition, we have $\{w_{ni}\}$ which are numerical approximations of the exact solution constructed at intermediate stages, $\tau_{ni}=t_n+c_i \Delta t$. We propose the following $s-$stage, explicit Runge-Kutta scheme for advancing the state from $t_n$ to $t_{n+1}$, for all $n\geq 0$,
\begin{subequations}\label{thescheme}
\begin{align}
w_{n1} &= w_{n},\\
w_{nj} &= \int_0^{\infty} H_n(k)e^{-c_j k^2} \: dk + \Delta t \sum_{i=1}^{j-1} a_{ji}(\Delta t) N_{\alpha}(w_{ni},\tau_{ni}), \quad 2 \leq j \leq s,\label{thescheme_wj} \\
w_{n+1} &= \int_0^{\infty} H_n(k) e^{-k^2} \: dk + \Delta t \sum_{i=1}^s b_i(\Delta t)  N_{\alpha}(w_{ni},\tau_{ni}),\label{thescheme_w}\\
H_{n+1}(k) &= e^{-k^2} H_{n}(k) + \frac{2}{\pi} \frac{\tilde{\gamma}}{\tilde{\gamma}^2 + k^2}\Delta t \sum_{i=1}^s d_i(\Delta t;k)  N_{\alpha}(w_{ni},\tau_{ni}), \label{thescheme_H}
\end{align}
\end{subequations}
where $[a_{ji}]$ is the $s\times s$ 
 \textit{Runge-Kutta matrix}, $\{b_i\}, \{d_i(k)\}$ are scheme-specific \textit{weights} and $\{0\leq c_i \leq 1\}$ are the\textit{ explicit Runge-Kutta nodes}. The initial conditions for the solver are:
\begin{equation}
[w_0, H_0(k)]^T = \left[w_0,\frac{2}{\pi}\frac{\tilde{\gamma}}{k^2+\tilde{\gamma}^2} w_0\right]^T~.
\end{equation}
The standard \textit{Butcher tableau} representation for the nodes and weights of the above constructed scheme is given as,
\begin{equation}\label{eqn:std_butcher}
\centering
\begin{tabular}{c | c c c c c c c}
$c_1=0$ &  & & & & & &\\
$c_2$ & $a_{21}$ & & & & & &\\
$c_3$ & $a_{31}$ & $a_{32}$ & & & & &\\
. & ... &  & & & & &  \\
$c _s$& $a_{s1}$ & $a_{s2}$ &  &...& & $a_{s,s-1}$ & \\
\hline
& $b_1$ & $b_2$ &   & ... & & &$b_s$ \\
& $d_1(k)$ & $d_2(k)$ &   & ... &  & & $d_s(k)$
\end{tabular}
\end{equation}
We demand $c_1=0$ and $a_{i,j\geq i}=0$ due to the requirement of constructing an explicit scheme. The above representation differs slightly from the standard representation in that there is an extra row of information for the functions $d_i$ corresponding to the weights for the evolution of the history function (Eq. (\ref{thescheme_H})). For a given user-defined choice of nodes $c_i$, an $s-$stage scheme has $s(s+3)/2$ unknown coefficients to be fixed. We introduce the following shorthand notations for the recurring functions used in defining the scheme-parameters -
\begin{equation}
\phi_{m}(\tilde{\gamma}) = \int_0^1 \chi(1-\tau;\tilde{\gamma}) \tau^m \: d\tau~, \quad \phi_{m,n}(\tilde{\gamma}) = \int_0^1 \chi(c_n(1-\tau); \tilde{\gamma}) \tau^m \: d\tau~, \quad \psi_{m}(k) = \int_0^1 e^{-k^2(1-\tau)} \tau^m \:d\tau ~,
\end{equation}
where $n$ is an integer index $(\geq 1)$ identifying the Runge-Kutta nodes, and $m$ is a positive real number. These parameters are a one-time computation once $\Delta t$ is specified. These can be pre-computed before time-iteration begins.

\subsection{Two-stage, first-order Runge-Kutta method}{\label{s:scheme2f}}
\noindent The two-stage, explicit Runge-Kutta method of global first-order accuracy is constructed by satisfying the following order conditions-
\begin{subequations}{\label{oc:2f}}
\begin{align}
\sum_{j=1}^2 c_j^{(i-1)/2} b_j &= \phi_{(i-1)/2}, \quad 1\leq i \leq 2~, \label{oc1:2f}\\
\sum_{j=1}^2 c_j^{(i-1)/2} d_j(k) &= \psi_{(i-1)/2}(k), \quad1\leq i \leq 2~, \label{oc2:2f}\\
a_{21} &= c_2 \phi_{0,2} \label{oc3:2f}~,
\end{align}
\end{subequations}
where the second node, $c_2$, is a free parameter leading to the following one-parameter family of schemes - 
\begin{equation}{\label{bt:2f}}
\begin{tabular}{ c | c c}
0 & & \\
$c_2$ & $c_2\phi_{0,2}$ &  \\  
\hline \\ \vspace{-20pt} & & \\ \vspace{-5pt}
 &$ \phi_0 - \frac{1}{c_2^{1/2}}\phi_{1/2} $& $\frac{1}{c_2^{1/2}}\phi_{1/2}$ \\ & & \\ 
& $\psi_0(k)  - \frac{1}{c_2^{1/2}} \psi_{1/2}(k) $& $\frac{1}{c_2^{1/2}}\psi_{1/2}(k)$
\end{tabular}
\end{equation}
\noindent We choose $c_2=1$ to obtain a first-order scheme where the global error scales linearly with $\Delta t$.

\subsection{Four-stage, second-order Runge-Kutta method}{\label{s:scheme4f}}
\noindent The four-stage Runge-Kutta scheme can be constructed analogous to the two-stage scheme by satisfying the following order conditions using four intermediate stages-
\begin{subequations}{\label{oc:4f}}
\begin{align}
\sum_{j=1}^4 c_j^{(i-1)/2} b_j &= \phi_{(i-1)/2}, \quad 1\leq i \leq 4, \label{oc1:4f}\\
\sum_{j=1}^4 c_j^{(i-1)/2} d_j(k) &= \psi_{(i-1)/2}(k), \quad 1\leq i \leq 4, \label{oc2:4f} \\
\sum_{i=1}^{j-1} a_{ji} &= c_j \phi_{0,j}, \quad 2\leq j \leq 3, \label{oc3:4f}\\
b_3 a_{32}c_2^{1/2} &= b_3 c_3^{3/2} \phi_{\frac{1}{2},3} + b_2 c_2^{3/2} \phi_{\frac{1}{2},2}, \label{oc4:4f}\\
\sum_{j=1}^3 a_{4j} c_j^{(i-1)/2} &= c_4^{(i+1)/2} \phi_{(i-1)/2,4}, \quad 1\leq i \leq 3 \label{oc5:4f}
\end{align}
\end{subequations}
where once again, besides setting $c_1=0$, the choice of rest of the nodes as free parameters leads to the following three-parameter family of schemes-
\begin{equation}{\label{bt:4f}}
\centering
\begin{tabular}{c | c c c c}
\centering
0 & & & \\
$c_2$ & $c_2 \phi_{0,2}$ &  &\\
$c_3$ & $c_3\phi_{0,3}-a_{32}$ & $\frac{b_3 c_3^{3/2}\phi_{\frac{1}{2},3}+b_2 c_2^{3/2} \phi_{\frac{1}{2},2}}{b_3 c_2^{1/2}}$ &  \\
$c_4$& $c_4 \phi_{0,4} - a_{42} - a_{43}$ & $\frac{c_4^2 \phi_{1,4}}{c_2} - \frac{c_3}{c_2} a_{43}$ & $\frac{c_2^{1/2} c_4^{3/2} \phi_{\frac{1}{2},4}-c_4^2 \phi_{1,4}}{(c_2 c_3)^{1/2}-c_3}$ &\\
\hline
& $b_1$ & $b_2$ & $b_3$ & $b_4$  \\
& $d_1(k)$ & $d_2(k)$ &  $d_3(k)$ & $d_4(k)$
\end{tabular}
\end{equation}
In the above, $b_i$'s, $d_i$'s are obtained by solving the matrix equations Eqs.\ref{oc1:4f}-\ref{oc2:4f} once a choice for the nodes $c_j$ is made. With the $b_i,d_i$ at hand, we compute $a_{ij}$ from right-to-left (see Eq. \ref{eqn:std_butcher} for Butcher tableau notation). It is important to choose the $c_j$'s carefully such that the matrices in Eq.\ref{oc:4f} are invertible. For instance, choosing any two nodes equal would render the associated matrices non-invertible. We use the following choice for the nodes- $(0,0.25,0.9,1)$. The estimated convergence rate is $\Delta t^{2}$. Higher order accurate schemes may be derived. 

\section{Derivation of the numerical scheme and error analysis}{\label{sec5}}
To derive the Runge-Kutta (RK) scheme given in Sections \ref{s:scheme2f} and \ref{s:scheme4f}, we follow the construction procedure of explicit exponential time differencing Runge-Kutta method given in \cite{HO2005}. Our construction of the scheme relies on the representation of the exact solution at arbitrary times in the following form (a generalization of Eq. \ref{newstatetransformed1})
\begin{equation}{\label{eq:vopI}}
    w(t_n+\theta \Delta t) = \int_0^{\infty} e^{-\theta k^2}H(k,t_n) \:dk + \int_0^{\theta \Delta t} \chi(\theta \Delta t-\tau;\tilde{\gamma})N_{\alpha}(w(t_n+\tau),t_n+\tau) \:d\tau~,
\end{equation}
where $\theta \in (0,1]$. We derive the RK-type quadrature formulae by approximating the time-integral to the desired accuracy. The additional distinguishing task is the computation of the history integral over $k$ of the form in Eq. (\ref{eq:vopI}). Our strategy is to ensure that the error due to approximation of the history integrals is subdominant to that due to approximation of the time-integrals (see Section \ref{subs5.2}). Thus when we derive the RK weights in Section \ref{subs5.1}  we retain the semi-infinite integral for $H(k,t_n)$ in the representation of the history integral. \\
\indent We state some properties of the nonlinear function which are used in deriving the RK quadrature. We seek an expansion for the nonlinear function $N_{\alpha}(w(t),t)$ in the neighbourhood of a time grid point $t=t_n$. We assume there is a function $\tilde{N}_{\alpha}$ such that  $\tilde{N}_{\alpha}(t)=N_{\alpha}(w(t),t)$. We know that the solution to the linear MRG $\chi(t;\tilde{\gamma})$ has a series expansion in powers of $t^{1/2}$ around $t=0$ (\textit{Property 1}), and a regular expansion in integer powers at later times. Using standard arguments one can show that this holds for the solution function $w$ as well (see Eq. \ref{MR:duhamel}). Consequently this property is inherited by $N_{\alpha}(w)$ too. This allows us to write the following expansion for $N_{\alpha}$
\begin{equation}{\label{seriesexp}}
    N_{\alpha}(w(t_n+\tau),t_n+\tau) = \tilde{N}_{\alpha}(t_n+\tau)=\sum_{m=0}^{\infty}C_m(t_n) \tau^{m/2}, \quad \forall n~,
\end{equation}
with coefficients $C_m$ that are functions of the `base point' $t_n$. Note that the above representation uniformly captures both the short-time behaviour as well as the long-time behaviour since at $t_n=0$ these coefficients are non-zero while at later-times the coefficients corresponding to the odd index $m$ become zero, resulting in a regular Taylor series expansion. 

\subsection{Computation of the RK matrix and weights}{\label{subs5.1}}
\indent The  state-variable whose error we want to control is the slip velocity $w$. Therefore we write the equation for the associated error. We introduce the primary error function $e_n = w_n-w(t_n)$ and the auxiliary error functions, $E_{ni} = w_{ni}-w(\tau_{ni})$ and $e_n^H(k)=H_n(k)-H(k,t_n)$, which denote the difference between exact state variables and their numerical counterparts. We subtract Eq. \ref{newstatetransformed1} from Eq. \ref{thescheme_w} to get the following master equation for the error in slip velocity at any $n$,
\begin{equation}{\label{eq:masterError}}
    e_{n+1} =\int_0^{\infty} e^{-k^2}e_n^H(k) \:dk + \Delta t\sum_{i=1}^{s}b_i\Big[N_{\alpha}(w_{ni},\tau_{ni})-N_{\alpha}(w(\tau_{ni}),\tau_{ni})\Big] +\delta_{n+1}~.
\end{equation}
 We delineate the different contributions to the primary error function: $(i)$ the defect term $\delta_{n+1}$ corresponds to error solely due to numerical approximation of the exact time-integral by the RK-type quadrature rule, $(ii)$ the terms proportional to $N(w_{ni},\tau_{ni})-N(w(\tau_{ni}),\tau_{ni})$ computed at the intermediate stages $\tau_{ni}$ correspond to error due to the difference in the nonlinear action on exact and the corresponding numerical approximation of the state variable, and $(iii)$ the integral term contains the error due to numerical approximation of the history function. Controlling these different sources of error results in the desired order of accuracy in the numerical scheme. We derive conditions called \textit{order conditions} from each contribution such that the numerical method \ref{thescheme} converges with order $p$, that is, the error function $e_n \sim \mathcal{O}(\Delta t^{p})$ as $\Delta t \rightarrow 0$.\\
\indent $(i)$ First we estimate the contribution from the defect term $\delta_{n+1}$ in terms of $\Delta t$. The RK weights $\{b_i\}$ are fixed by controlling this defect term. We evaluate the numerical scheme Eq. \ref{thescheme_w} at the exact solution, that is, we substitute $w_{ni}= w(\tau_{ni}), w_n = w(t_n), H_n(k) = H(k,t_n)$, yielding,
\begin{equation}
 w(t_{n+1}) = \int_0^{\infty} H(k,t_n) e^{-k^2} \:dk + \Delta t \sum_{i=1}^s b_i N_{\alpha}(w(\tau_{ni}),\tau_{ni}) -\delta_{n+1}
\end{equation}
where $\delta_{n+1}$ is the defect term reflecting error due to the RK-quadrature. We subtract the above from the exact expression for $w(t_{n+1})$ in Eq. \ref{newstatetransformed1} and use the series expansion in Eq. \ref{seriesexp} to get the following expression for the defect term:
\begin{equation}
\begin{aligned}
\delta_{n+1} &= \Delta t \sum_{i=1}^s b_i N_{\alpha}(w(\tau_{ni}),\tau_{ni}) - \int_0^{\Delta t} \chi(\Delta t-\tau;\tilde{\gamma})N_{\alpha}(w(t_n+\tau),t_n+\tau) \:d\tau \\
&=\sum_{m=1}^{p_1} C_{m-1}(t_n)\Delta t^{\frac{m+1}{2}}\Bigg[\sum_{i=1}^{s} b_i c_i^{(m-1)/2} -\phi_{(m-1)/2}(\tilde{\gamma})\Bigg] + \mathcal{O}(\Delta t^{{p_1}/2 +1 })~,
\end{aligned}
\end{equation}
where $p_1$ is a positive integer indicating the extent of truncation. The expression in the parenthesis parameterized by $m$ are generators of $p_1$ number of the order conditions. For the schemes we have derived in Sections \ref{s:scheme2f} and  \ref{s:scheme4f}, we have set $p_1 = s$. The order conditions \ref{oc1:2f} and \ref{oc1:4f} result from setting these expressions for $m\leq p_1=s$ to zero. As a result across the two schemes, we have
\begin{equation}
    \delta_{n+1} \sim \mathcal{O}(\Delta t ^{s/2+1})~.
\end{equation}
\indent $(ii)$ Next we estimate the contribution due to the nonlinear function, $N_{\alpha}$, in the master equation. The Runge-Kutta matrix, $[a_{ij}]$, is fixed by controlling this contribution.  We recall the stage error function, $E_{ni} = w_{ni} - w(\tau_{ni})$, and state the following relation (\ref{A2}):
\begin{equation}{\label{eq:Nexpansion}}
N^{\text{diff}}_{n,i}:=N_{\alpha}(w_{ni},\tau_{ni}) - N_{\alpha}(w(\tau_{ni}),\tau_{ni}) = J_n E_{ni} + \mathcal{O}(E_{ni} \sqrt{\Delta t})
\end{equation}
where $J_n = \partial N_{\alpha}/\partial w$ at $t=t_n$. Substituting this in the second term on right-hand side in Eq.\ref{eq:masterError} results in the leading order contribution from the term $\Delta t \sum_{i=1}^s b_i J_n E_{ni}$. To make sense of this in terms of $\Delta t$ we further need to estimate the errors at the intermediate stages, $E_{ni}$. In order to do that, we write the equation for the stage error function obtained by subtracting exact expression (\ref{eq:vopI} for $\theta=c_j$) from the numerical expression (\ref{thescheme_wj}) evaluated at the intermediate stage, $\tau_{nj}$:
\begin{equation}\label{eqn:Enj}
\begin{aligned}
E_{nj} &= \int_0^{\infty} e^{-c_jk^2} e^H_n(k) \: dk + \Delta t \sum_{i=1}^{j-1} a_{ji}N^{\text{diff}}_{n,i} + \delta_{nj},  \quad j > 1\\
&= \int_0^{\infty} e^{-c_jk^2} e^H_n(k) \: dk + \Delta t \sum_{i=1}^{j-1} a_{ji}(J_n E_{ni}+\mathcal{O}(E_{ni} \sqrt{\Delta t})) + \delta_{nj},
\end{aligned}
\end{equation}
where $\delta_{nj}$ denotes the defect due to RK quadrature at the intermediate stage. Note that $E_{n1}=e_n$, $\delta_{n1}=0$. Order analysis of the stage errors, $E_{nj}$, follows similarly to Eq. \ref{eq:masterError} due to identical decomposition of the error contributions. Also note that due to recursive construction, the second term on the right in Eq. \ref{eqn:Enj} is subdominant to the rest of the terms ($\Delta t$ times the rest of the terms). Thus, it is sufficient to check the contribution of $\delta_{nj}$ and $e_n^H$ to the stage error $E_{nj}$. \\
\indent We examine the term $\Delta t \sum_{i=1}^s b_i J_n \delta_{ni}$. An expression for $\delta_{nj}$ (for $j>1$) can be obtained by repeating the procedure followed for estimating $\delta_n$:
\begin{equation}
\delta_{nj} = \sum_{m=1}^{p_{2,j} }C_{m-1}(t_n)\Delta t^{\frac{m+1}{2}}\Bigg[\sum_{i=1}^{j-1} a_{ji} c_i^{(m-1)/2}-c_j^{(m+1)/2} \phi_{(m-1)/2,j}(\tilde{\gamma})\Bigg] + \mathcal{O}(\Delta t^{p_{2,j}/2+1})~,
\end{equation}
where $p_{2,j}$ are positive integers. Using the above expression, we collect terms of different powers of $\Delta t$ in the expression $\Delta t \sum_{i=1}^s b_i J_n \delta_{ni}$ and set the coefficients to zero up to the desired order. For the two-stage scheme in Section \ref{s:scheme2f}, we set $p_{2,2} = 1$, and for the four-stage scheme in Section \ref{s:scheme4f}, we set $p_{2,2}=2=p_{2,3}$ and $p_{2,4}=3$. These generate the order conditions \ref{oc3:2f} for the 2-stage scheme and \ref{oc3:4f}, \ref{oc4:4f}, and \ref{oc5:4f} for the 4-stage scheme. The resultant contribution can be shown to be 
\begin{equation}
     \Delta t\sum_{i=1}^{s}b_i(N_{\alpha}(w_{ni},\tau_{ni})-N_{\alpha}(w(\tau_{ni}),\tau_{ni}) \sim \mathcal{O}(e_n^H(k)\Delta t )+\mathcal{O}(\Delta t^{s/2 +1})~.
\end{equation}
\indent $(iii)$ Finally, we estimate the contribution of error due to the numerical approximation of history integral. This fixes the weight functions, $\{d_i(k)\}$. Once again, we write the equation for the error corresponding to the history function, $e_n^H(k) = H_n(k)-H(k,t_n)$, by subtracting Eq. \ref{newstatetransformed2} from Eq. \ref{thescheme_H}  :
\begin{equation}{\label{eq:errorH}}
    e^H_{n+1}(k) = e^{-k^2} e_n^H(k) + \frac{2}{\pi} \frac{\tilde{\gamma}}{k^2 + \tilde{\gamma}^2}\Delta t \sum_{i=1}^s d_i(k)N^{\text{diff}}_{n,i} + \delta_{n+1}^H(k)~.
\end{equation}
We begin by estimating the history-defect term, $\delta_{n+1}^H$, whose expression is derived by subtracting Eq. \ref{thescheme_H} from Eq. \ref{newstatetransformed2} evaluated at the exact state variables:
\begin{equation}\label{eq:delta_n1_H}
    \delta_{n+1}^H(k) = \sum_{m=1}^{p_3} C_{m-1}(t_n)\Delta t^{\frac{m+1}{2}}\Bigg[\sum_{i=1}^{s} d_i(k) c_i^{(m-1)/2} -\psi_{(m-1)/2}(k) \Bigg] + \mathcal{O}(\Delta t^{{p_3}/2 +1 })~,
\end{equation}
where $p_3$ is a positive integer. Setting the expression in the parenthesis to zero for $m\leq p_3 = s$ generates as many order conditions as the number of unknown functions $d_i(k)$ for every $k$. Therefore we can solve for $d_i(k)$'s  (see Eq. \ref{oc2:2f} and Eq. \ref{oc2:4f}). At this point we have fixed all the unknown parameters in our RK scheme namely, $a_{ij}, b_i, d_i(k)$. Demanding that any further conditions be satisfied would result in contradiction from the already derived order conditions. Thus the dependence of the history error function, $e_{n+1}^H$, on $\Delta t$ can no further be controlled by construction. As a result, we analyze the behaviour of Eq. \ref{eq:errorH} under the constraints of the derived order conditions.

Let $\rho(k) := (2/\pi)(\tilde{\gamma}/(k^2+\tilde{\gamma}^2))$. Using the history error function in Eq. \ref{eq:errorH} recursively, we get
\begin{equation}{\label{eq:recurErrorH}}
    e_{n+1}^H(k) = e^{-(n+1)k^2} e_0^H(k) + \rho(k) \Delta t\sum_{m=0}^n \sum_{i=1}^s d_i(k) e^{-mk^2} N^{\text{diff}}_{n-m,i}+ \sum_{m=0}^n e^{-mk^2} \delta_{n+1-m}^H(k)~,
\end{equation}
where the first term has no contribution since $e_0^H(k)=0$ by imposition of the exact initial condition. Further due to Eq. \ref{eq:delta_n1_H} and $(n+1)\Delta t = t_{n+1}$, the last term is
\begin{equation}
    \sum_{m=0}^n e^{-mk^2} \delta_{n+1-m}^H(k) \sim \mathcal{O}(\Delta t^{s/2}).
\end{equation}
\noindent Using Eq. \ref{eq:Nexpansion} and using the same argument as before that the major contribution to $E_{ni}$ in $\Delta t \sum_{i=1}^s d_i(k)J_n E_{ni}$ comes from $e_n^H(k)$ and $\delta_{ni}$, the second term in Eq. \ref{eq:recurErrorH} can be shown to be (see \ref{A3} for contribution from $\delta_{ni}$):
\begin{equation}{\label{showninC}}
\begin{aligned}
    \Delta t\sum_{m=0}^n \sum_{i=1}^s d_i(k) e^{-mk^2} N^{\text{diff}}_{n-m,i} & \sim \Delta t\sum_{m=0}^n \sum_{i=1}^s d_i(k) e^{-mk^2} J_{n-m} E_{n-m,i}\\
    & \sim \Delta t\sum_{m=0}^n \sum_{i=1}^s d_i(k) e^{-(n-m)k^2} J_{m} \int_0^{\infty} e^{-c_i q^2}e_m^H(q)\:dq + \mathcal{O}(\Delta t^{s/2})~.
\end{aligned}
\end{equation}
\noindent Combining the above estimates, Eq. \ref{eq:recurErrorH} reduces to
\begin{equation}
    e_{n+1}^H(k) = \rho(k)\Delta t\sum_{i=1}^s d_i(k) \Bigg[\int_0^{\infty} e^{-c_i q^2} \Big(  \sum_{m=0}^n e^{-k^2(n-m)} J_m e_m^H(q)\Big) \:dq\Bigg] + \mathcal{O}(\Delta t^{s/2})~.
\end{equation}
Using mathematical induction for different $n$, we can show the following
\begin{equation}
    e_{n+1}^H(k) \sim \mathcal{O}(\Delta t^{s/2}) \implies \int_0^{\infty} e^{-k^2} e_n^H(k) \:dk \sim \mathcal{O}(\Delta t^{s/2}) ~.
\end{equation}
As a result of the contributions discussed above, we have
$e_{n+1} \sim \mathcal{O}(\Delta t^{s/2})$.
\begin{table}[tbh!]
    \centering
    \begin{tabular}{c|c|c c|c|c}
    \hline
         $s$ & $p_1$  & $j$ &$p_{2,j}$& $p_3$ & $p$\\
         \hhline{=|=|==|=|=}
         2 & 2 & $2$ & 1 & 2 & 1\\
         \hline
          &  & 2 & 2 & &\\
         4 & 4 & 3 & 2 & 4 & 2\\
          &  & 4 & 3 & &\\
          \hline
    \end{tabular}
    \caption{Summary of parameters $(p_1, \{p_{2,j}\},p_3)$  used to control the defect terms $(\delta_n, \{\delta_{nj}\}, \delta_n^H$) in two-stage $(s=2)$ and four-stage $(s=4)$ schemes to get $(\Delta t)^p$-accurate Runge-Kutta scheme (see Section \ref{sec4}).}
    \label{tab:my_label}
\end{table}
\subsection{Computation of the history integral}{\label{subs5.2}}
\indent We use the Clenshaw-Curtis quadrature rule to approximate the history integrals in the scheme (Eq. \ref{thescheme}). It is a common quadrature rule used to approximate integrals over infinite and semi-infinite lengths.  As a pre-requisite, we  map the semi-infinite interval $k \in [0,\infty)$ to $\tilde{k}\in [-1,1]$ under the transformation $k = \sqrt{\tilde{\gamma}}(1+\tilde{k})/(1-\tilde{k})$. We expand the resulting integrand in the basis of Chebyshev polynomials of the first kind, $\{T_m\}$.  In summary, 
\begin{equation*}
    \int_0^{\infty} e^{-k^2} H_n(k) \:dk = \int_{-1}^{1} e^{-k(\tilde{k})^2}H_n(k(\tilde{k})) \frac{dk}{d\tilde{k}} \: d\tilde{k}=: \int_{-1}^1 F_n(\tilde{k})\:d\tilde{k}~,
\end{equation*}
where we have compactly defined the integrand  as $F_n$. The Clenshaw-Curtis quadrature rule effectively replaces the integral with the following finite sum
\begin{equation}
    \int_{-1}^1 F_n(\tilde{k})\:d\tilde{k} \approx \sum_{m=1}^{M+1} \mu_m F_n(\tilde{k}_m),
\end{equation}
where $\tilde{k}_m = \cos(\theta_m)$ are the Chebyshev nodes computed at linearly-spaced $\theta_m =  \pi(m-1)/M$, and $\mu_m$ are the associated weights. This procedure yields spectral accuracy with respect to the number of quadrature points, $M+1$. It is important to note that this quadrature is a recurring operation to be performed at every time-iterate and its subsuming stages since the integrand is a function of a dynamically evolving state-variable, $H_n$. However, this is a time-independent cost once $M$ is fixed at the beginning of the simulation.\\
\indent The RK weights were derived in the previous section assuming that there is no quadrature error incurred in computing the history-error integral in the primary error function (first term on the right-hand side in Eq. \ref{eq:masterError}). But we have used the Clenshaw-Curtis quadrature as described above. We show that it is justified even as the quadrature error is consistent up to the same order of accuracy as the time-convolution RK quadrature. We make an estimate for the error associated with the numerical approximation of the history integral term in the scheme. We define an associated error quantity, $\zeta_n$
\begin{equation*}
    \zeta_n := \sum_{m=1}^{M+1} \mu_m F_n(\tilde{k}_m)-\int_0^{\infty} e^{-k^2} H(k,t_n) \:dk.
\end{equation*}
We add and subtract the appropriate quantity to get the following decomposition for error contributions 
\begin{equation*}
\begin{aligned}
       \zeta_n &= \Big(\int_0^{\infty} e^{-k^2} H_n(k)\:dk-\int_0^{\infty}e^{-k^2}H(k,t_n) \:dk \Big) + \Big(\sum_{m=1}^{M+1} \mu_m F_n(\tilde{k}_m)-\int_0^{\infty} e^{-k^2} H_n(k)\:dk \Big),\\
       & = \int_0^{\infty} e^{-k^2} e_n^H(k) \:dk + \zeta_n^{Q}~,
\end{aligned}
\end{equation*}
where $\zeta_n^Q$ is the error solely due to the quadrature approximation of the history integral (in our case the Clenshaw-Curtis quadrature). In principle any quadrature rule can be used and the associated convergence property will be reflected in $\zeta_n^Q$. Hence this justifies the assumption we made in the beginning of the section that this quantity virtually sets the ``machine precision'' for the scheme. Note that the remaining integral term with the history error function is what appears in the master error Eq. \ref{eq:masterError}. 

\section{Numerical experiments}{\label{sec6}}
In this section, we test the error bounds for the first- and second-order Runge-Kutta schemes for the MRG equation described in Sections \ref{s:scheme2f} and \ref{s:scheme4f} for two model problems.
\subsection{Oscillating-in-time force}
\label{sec6p1}
\noindent We consider the following 1D scalar MRG equation with a time-oscillating force ($N = \sin(\omega t))$:
\begin{equation}
    \frac{dw}{dt} + \gamma\Big(\frac{w_0}{\sqrt{\pi t}} + \int_0^{t} \frac{d w/d\tau}{\sqrt{\pi(t-\tau)}} \: d\tau \Big)= -\alpha w +\sin(\omega t)~,
\end{equation}
for $t\in (0,T]$, subject to the initial condition $w(0)=w_0$. This equation has the following closed-form analytical solution expression (Section 4.3 in \cite{sgp2019})
\begin{equation}{\label{exactsinsol}}
    w(t) = \chi(t;\alpha,\gamma) w_0 + \frac{2\gamma}{\pi} \int_{0}^{\infty} \frac{k^2}{(-k^2 + \alpha)^2+k^2\gamma^2} \Bigg(\frac{k^2 \sin(\omega t)-\omega \cos(\omega t)}{k^4+\omega^2}+\frac{\omega e^{-k^2 t}}{k^4+\omega^2} \Bigg) \:dk~,
\end{equation}
where the semi-infinite $k-$ integral may be computed numerically using any available standard high-accuracy quadrature package. We used MATLAB's built-in numerical integrator function \textit{integrate} to generate the exact solution data.
\begin{figure}[!ht]
    \centering
    \subfloat[\centering ]{{\includegraphics[width=0.37\textwidth]{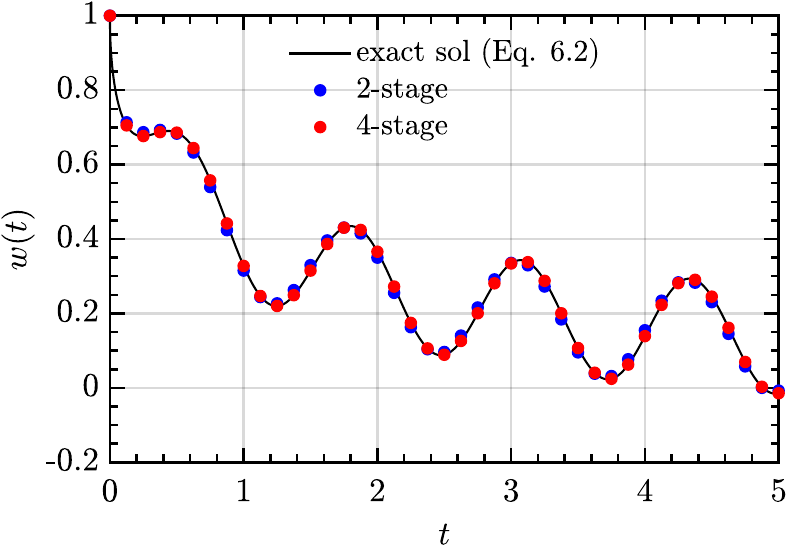} }}%
    \qquad
    \subfloat[\centering]{{\includegraphics[width=0.38\textwidth]{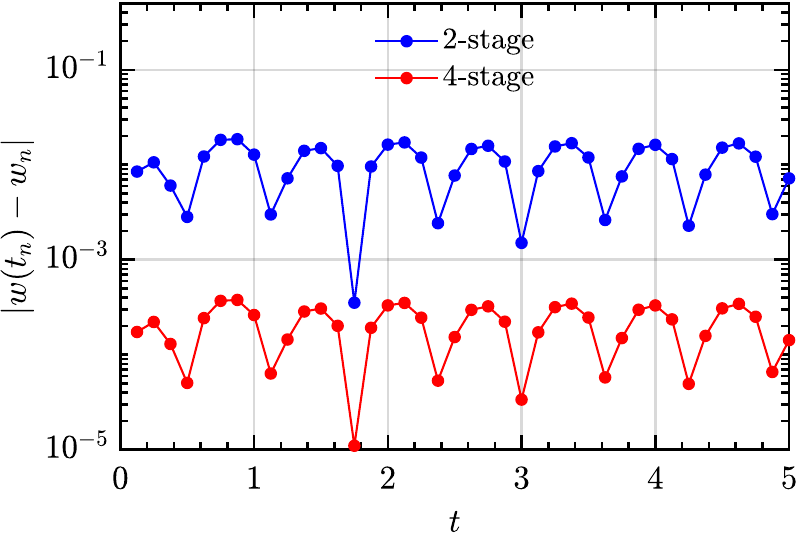}}}%
    \qquad
    \subfloat[\centering \label{fig:sinusoid:err_v_dt}]{{\includegraphics[width=0.39\textwidth]{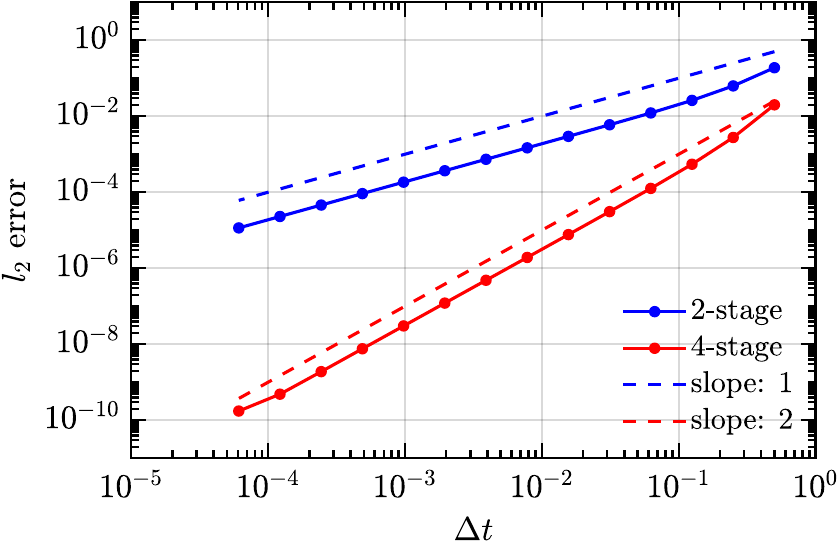} }}%
    \qquad
    \subfloat[\centering \label{fig:sinusoid:err_v_ctime}]{{\includegraphics[width=0.38\textwidth]{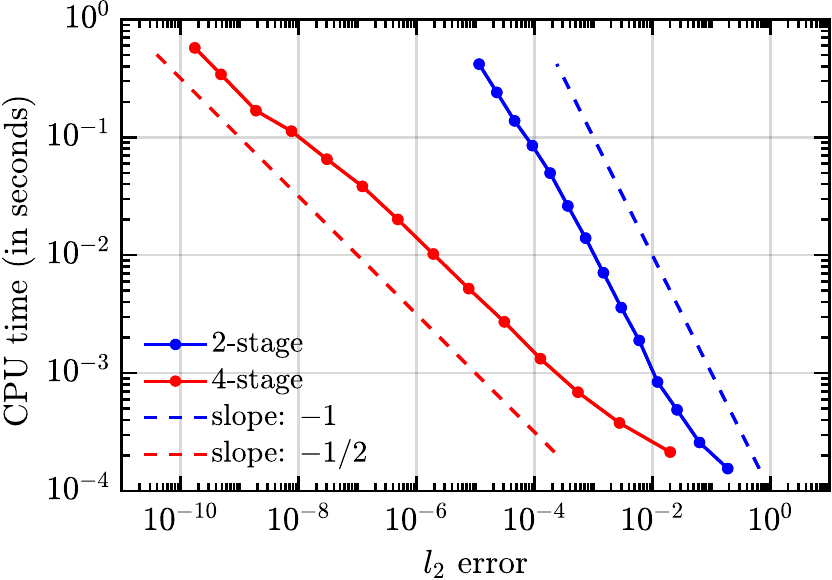} }}%
    \caption{Numerical solution of 1D scalar MRG equation forced with $N = \sin(\omega t)$ and subject to initial condition, $w_0=1$, for the parameters $(\alpha,\gamma,\omega,T) = (0.33,1,5,5)$ using the 2- and 4-stage schemes (\ref{oc:2f}  and \ref{oc:4f}). $M=51$ (Chebyshev) quadrature points were used to compute the history integral. (a) Comparison of the exact solution with the numerical solutions for $\Delta t = 2^{-3}$. (b) Error as a function of time for $\Delta t = 2^{-3}$. (c) Scaling of error with time-step $\Delta t$. The errors are measured against the analytical solution (\ref{exactsinsol}) in the $l_2-$norm. The slopes are consistent with the estimated orders of accuracy of schemes. (d) Scaling of operational cost with error. CPU time is used as a proxy for the operational cost. The slopes of the curves indicate the scheme's cost to improve in accuracy. As per the scheme's algorithm, $l_2$ error $\sim \Delta t^p$, where $p=\{1,2\}$ is the order of accuracy, and operational cost $\sim \Delta t^{-1}$. The slopes of the curves verify this estimate. The codes for the schemes used in this example are available on GitHub at \url{https://github.com/jagannathan-divya/rk4mrg}.}
    \label{fig:sinusoid}%
\end{figure}
We numerically evolve the model equation over discrete times $t_n = n\Delta t$ using the two schemes derived in Section \ref{sec4} for a chosen set of parameters. According to the construction of the scheme, we expect the two- and four-stage schemes to generate first- and second-order rates of convergence for global error respectively. This is verified in Figure \ref{fig:sinusoid:err_v_dt}, where the error is measured in the $l_2$ norm 
\begin{equation}
    l_2 \text{ error} =  \Big(\Delta t\sum_{n=1}^{N} |w_n-w(t_n)|^2\Big)^{1/2}~.
\end{equation}
Additionally, in Figure \ref{fig:sinusoid:err_v_ctime}, we measure the computational effort required to achieve a desired accuracy. We use CPU time taken to run the simulation up to $T=N\Delta t$ as the proxy to estimate the operational cost (FLOPs) incurred by the scheme and plot it against the global error. In the non-Markovian treatment of the equation in the original formalism (\ref{Eq:compactMRG}), the CPU time scales quadratically with number of time-iterates, $N$, equivalently, $\Delta t^{-2}$. In contrast, in our Markovian reformulation, we recover the linear scaling for a canonical dynamical system (equivalently, $\Delta t^{-1}$).

\subsection{Particle in 2D Lamb-Oseen vortex}
We next consider a vector example of a particle moving according to the MRG equation in a 2D stationary flow field  where the vector components are coupled via the flow. This problem, in addition to the vector equations for the particle slip velocity $\textbf{w}$, requires the evolution of the particle position vector $\textbf{y}$ to determine the instantaneous nonlinear forcing of the flow at the particle location. The system is
\begin{subequations}
    \begin{align}
       \frac{d{\textbf{y}}}{dt} &= \textbf{w} + \textbf{u}(\textbf{y}(t)),\\
        \frac{d\textbf{w}}{dt} + \gamma \Big(\frac{\textbf{w}_0}{\sqrt{\pi t}} + \int_0^{t} \frac{d\textbf{w}/d\tau}{\sqrt{\pi(t-\tau)}} \: d\tau \Big) &= -\alpha\textbf{w} + \Big(\frac{1}{R}-1\Big)\frac{D\textbf{u}}{Dt}(\textbf{y}(t)) - \textbf{w}\cdot\nabla \textbf{u}(\textbf{y}(t)),
    \end{align}
\end{subequations}
where $t\in (0,T]$, subjected to initial conditions $\textbf{y}(0)=\textbf{y}_0, \textbf{w}(0)=\textbf{w}_0$, and a 2D stationary Lamb-Oseen vortex, $\textbf{u}(\textbf{r})=(1-e^{-|\textbf{r}|^2})/|\textbf{r}| \hat{e}_{\theta}$ for the background flow.
\begin{figure}[!ht]
    \centering
    \subfloat[\centering ]{{\includegraphics[width=0.3\textwidth]{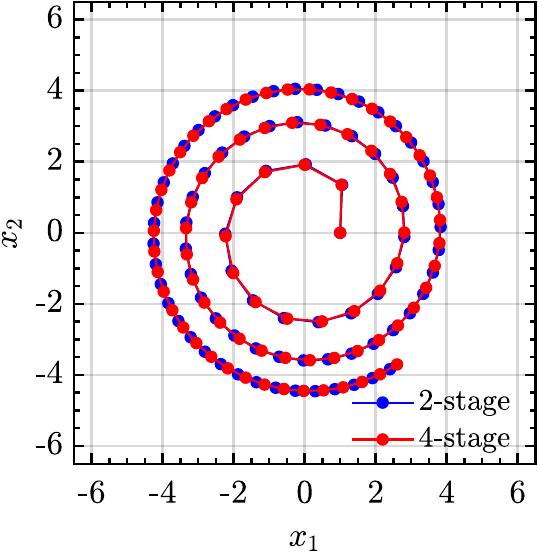} }}%
    \qquad
    \subfloat[\centering]{{\includegraphics[width=0.475\textwidth]{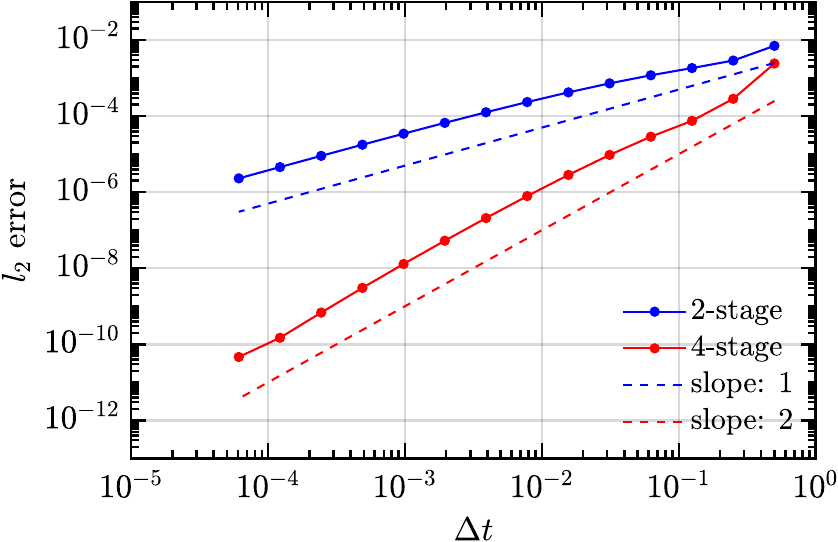}}}%
    \caption{Numerical advection of a particle in 2D stationary Lamb-Oseen vortex starting at $\textbf{y}_0=(1,0)$, with non-zero initial slip velocity $\textbf{w}_0=(1,0)$, for the parameters $(\alpha,\gamma) = (1,1)$ using our 2-stage and 4-stage schemes. $M=51$ (Chebyshev) quadrature points were used to compute the history integral. (a) Trajectory of the particle evolved up to $T=200$ with $\Delta t=2^{-3}$. (b) Scaling of error with time-step $\Delta t$ for simulations run up to $T=5$. Here, the errors are defined for the magnitude of slip velocity vector, $|\textbf{w}|$, and measured in the $l_2-$norm against a fine-resolution numerical solution computed using our 4-stage scheme with $\Delta t = 2^{-16}$ and $M=101$.  The codes for the schemes used in this example are available on GitHub at \url{https://github.com/jagannathan-divya/rk4mrg}.}%
    \label{fig:lamb_oseen}
\end{figure}
Due to lack of an analytical solution expression for the slip velocity, we treat the numerical solution obtained at a finer grid resolution to be the true solution. We then define error in computing the slip velocity as the departure from the fine grid solution. Figure \ref{fig:lamb_oseen} verifies the expected error convergence rates for the two schemes \ref{oc:2f} and \ref{oc:4f}.

\subsection{An alternative Markovian embedding procedure}{\label{sec6p3}}
An alternative natural way to construct the Markovian embedding was described in the overview Section \ref{sec:overview}.
We revisit this construction and compare it to the construction described in Section \ref{sec3} to demonstrate the computational advantage of the latter. 

Following Remark \ref{twoHs} and on comparing the definition of $w(t)$ in Eq. \ref{eq:Heqn} to that in Eq. \ref{newstatedynamics1}, one may identify the following to be the case for \ref{eq:Heqn}:
\begin{align} 
w(t_{n+1}) = \mathcal{R}^{\chi}_{\text{history}}(t_{n+1};t_{n+1}) = \int_{-\infty}^{\infty} H(k,t_{n+1}) \:dk\label{eqn:w-H:relation:no-discount}
\end{align}
with the same evolution equation for $H$ as in Eq. \ref{dyneq4H}. Note that the integrand here which is the history function itself, $H(k,t)$, doesn't have the decay property (in its $k$-dependence) that the integrand with the exponentially-discounted history function, $e^{-k^2 \Delta t} H(k,t)$, in Eq. \ref{newstatedynamics1} has. This suggests that a quadrature approximation of the integral in Eq. \ref{eqn:w-H:relation:no-discount} will require a larger number of basis functions. It is worth emphasizing that we are considering two different functions $H$, that satisfy the same differential equation, but differ in their relation to the slip velocity $w(t)$.

\begin{figure}[h!]
    \centering    \includegraphics[width=0.45\textwidth]{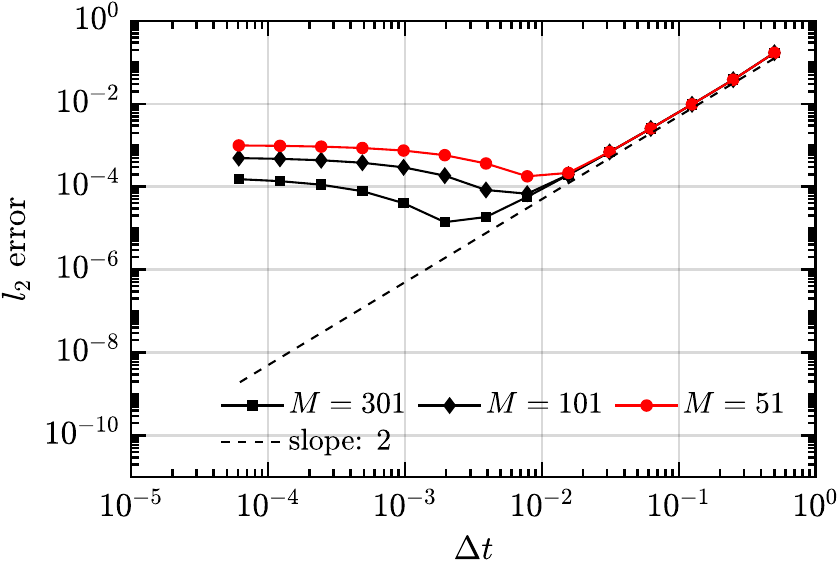}
    \caption{Repeat of numerical experiment \ref{sec6p1} under the \textit{alternative} Markovian embedding procedure (Section \ref{sec6p3}) for parameters $(\alpha,\gamma,\omega,T)=(0.33,1,5,5)$ using ETD2RK (Eq. 81, 82 in \cite{CM2002}). $M$ is the number of basis Hermite functions used to compute the history integral. Compare the red curve here with the red curve in Fig \ref{fig:sinusoid:err_v_dt} both of which correspond to $M=51$.  The codes used for the example are available on GitHub at \url{https://github.com/jagannathan-divya/rk4mrg}.}
    \label{fig4}
\end{figure}

Once again, we introduce the following re-definitions,
\begin{equation}{\label{alternateRedef}}
    \text{Re}(H(k,t)) \rightarrow H(k,t), \quad N_{\alpha}(w) = N(w) - \alpha w~,
\end{equation}
where the first reassignment here is due to
$\int \text{Im}(H(k,t)) \:dk = 0$ (see Eq. \ref{Hdef}). We then consider the expansion of the history function in the basis of Hermite functions, $\{\psi_m\}$,
\[ H(k,t) = \sum_{m=0}^{M} \mu_{2m}(t) \psi_{2m}(k)~,\]
where $\mu$ is an $(M+1)\times 1$ vector of time-dependent weights corresponding to the even-indexed Hermite functions. The odd-indexed Hermite functions do not contribute to the integral in (\ref{eqn:w-H:relation:no-discount}) due to their odd-symmetry. Substituting the Hermite expansion into Eq. \ref{dyneq4H} with the re-definitions \ref{alternateRedef}, and taking suitable inner-products, we derive the following evolution rule for the Hermite coefficient-vector in the Duhamel form:
    \begin{equation}{\label{evolveMu}}
        \mu(t+\Delta t) = e^{-L\Delta t } \mu(t) +f \int_0^{\Delta t} e^{-L(\Delta t-\tau)}  N_{\alpha}(w(t+\tau),t+\tau) \:d\tau,
    \end{equation}
where $L_{mn}=\int k^2 \psi_m(k)\psi_n(k) \:dk$ is an $(M+1)\times (M+1)$ tridiagonal matrix, and $f_{2m} = \gamma/\pi\int \psi_{2m}(k)/(k^2+\gamma^2) \:dk $ is an $(M+1)\times 1$ vector. The exponential matrix, $e^{-L\Delta t}$, is $Ve^{-\Lambda \Delta t}V^{-1}$ where $\Lambda$ is the diagonal matrix with eigenvalues of $L$ and $V$ is the corresponding eigenvector matrix. The resultant Eq. \ref{evolveMu} can be numerically solved using the exponential time differencing method of \cite{CM2002}. The slip velocity can then be reconstructed by:
\[  w(t+\Delta t) = \sqrt{2\pi} \sum_{m=0}^{M} \mu_{2m}(t+\Delta t)(-1)^m \psi_{2m}(0)~.\]
We repeat the numerical experiment in Section \ref{sec6p1} for the above construction using the second-order accurate ETD2RK scheme (\cite{CM2002}). In Figure \ref{fig4}, we plot the rate of error convergence by varying the number of basis functions used. We note that upon increasing the number of basis functions, the error in slip velocity converges with the rate $\mathcal{O}(\Delta t^2)$ over a range of $\Delta t$ that only slowly widens before eventually saturating. In comparison, the 4-stage, second-order scheme derived for the Markovian embedding procedure in Eq. \ref{newstatedynamics} required significantly fewer quadrature points to maintain the second-order accuracy over a much larger range of $\Delta t$ (See Fig. \ref{fig:sinusoid:err_v_dt}).

\section{Conclusions and discussion}{\label{sec7}}
We have derived explicit integration schemes for the MRG equation via the introduction of a co-evolving state variable which we call the history function. We have further established their formal rate of convergence and presented some numerical experiments to verify the same. Both the introduction of the co-evolving state variable and the derivation of the numerical integration scheme relied crucially on the spectral representation of the solution to the linear equation. The ideas presented here generalise to any equation with memory effects which can be expressed in the form
\[w(t) = \mathcal T(t)w_0 + \int_0^t \mathcal T(t-\tau)N(w(\tau),\tau)\: d\tau\]
where the linear solution operator $\mathcal T$ has a spectral representation,
\begin{equation}
    \mathcal{T}(t) = \int_{\Gamma} e^{f(k) t} g(k) \:dk~.\label{spec_rep}
\end{equation}
This form includes many FDEs where the non-local term appears linearly, including notably the MRG equation. The form of the integrand in Eq. \ref{spec_rep} determines the evolution equation for the co-evolving history function $H(k,t)$,
\begin{align}        
        \frac{dH(k,t)}{dt} &= f(k) H(k,t) + g(k) N(w(t),t) \label{genHevolve}~,
\end{align}
whereas the smoothness in $t$ for the function $\T$ is what determines the kind of expansions (integer or fractional powers of $t$) involved in the derivation of the RK scheme. The resultant numerical methods are explicit in time, have a memory cost that remains constant throughout the simulation, do not introduce any arbitrary design parameters (besides the step-size $\Delta t$) and can be generalised to obtain higher-orders of accuracy. Typical numerical methods for equations with memory effects have computational costs that grow quadratically with simulation time, whereas our methods have the linear growth seen in equations without memory-effects. This behaviour is due to the Markovian embedding suggested by the spectral representation. Finally, the simulation can be restarted from an arbitrary state given by the state variable $w$ and the history function $H(k,t)$ (or its numerical approximation). 

As seen in the MRG case, the construction of the numerical scheme for the above generalised system would entail similar computations, namely $(i)$ an appropriate quadrature rule for the history/residual integral $\mathcal R_{\text{history}}$. Often the decay properties of the integrand in Eq. \ref{spec_rep} will permit suitable deformation of the contour $\Gamma$ to the steepest descent contour where the cost of performing the quadrature will in principle be minimized; $(ii)$ a standard time-integrator to solve Eq. \ref{genHevolve} along with the original evolution equation.

A comment on the use of the history function in the place of non-local expression is in order. At first sight, it seems that both involve variables in infinite dimensional space. However, the history function makes an impossible calculation viable as higher modes in its wavenumber space lead to squared-exponentially smaller contributions, so a truncation is possible. This feature is not unique to the MRG equation, but will be true of any physical system where viscous and other dissipative effects will erase out higher spatial wavenumbers. In any particulate flow of fluids, for example in atmospheric clouds, the number of particles is so large that a continuously increasing memory requirement would be difficult to accommodate, whereas our local representation makes it easier to contemplate.

Memory effects in evolution equations can appear in combination with nonlinearity. Indeed this is the case for the evolution of moving boundary in the classical Stefan problem for phase change. Another notable example is the generalisation of the MRG equation for finite Reynolds number presented in \cite{LO1993}. These kinds of equations are not directly amenable to the method introduced in this paper, though the present method can be suitably adapted to such cases as well. The extension of the present method to problems with nonlinear non-local effects will be considered in an upcoming work.

\section*{Acknowledgements}
The numerical calculations reported in the paper were performed on the Mario
computing cluster at ICTS-TIFR. Research at ICTS-TIFR is supported by the Department of Atomic Energy Government of India, under Project Identification No. RTI4001. DJ acknowledges the hospitality provided at KITP, Santa Barbara during the program \textit{Multiphase flows in Geophysics and the environment (2022)} which was supported in part by the National Science Foundation under Grant No. NSF PHY-1748958, where part of this work was done. VV acknowledges support through the SERB MATRICS Grant (MTR/2019/000609) from the Science and Engineering Research Board (SERB), Department of Science and Technology, Government of India. 

\appendix
\section{Spectral representation of \texorpdfstring{$\chi(t;\alpha,\gamma)$}{} using the inverse Laplace transform}
\label{A1}
\noindent We have
\begin{equation}
    \chi(t;\alpha,\gamma) = \mathcal{L}^{-1}\Big[\frac{1}{s+\gamma \sqrt{s} + \alpha}\Big](t) = \frac{1}{2\pi i} \int_{a-i\infty}^{a+i\infty} \frac{e^{st}}{s+\gamma \sqrt{s} + \alpha} \:ds
\end{equation}
where the complex contour $(a-i\infty, a+i\infty)$ is the standard Bromwich's vertical contour such that all the singularities of the function $1/(s+\gamma \sqrt{s}+\alpha)$ lie to the left of it. We use the following parametric representation for complex $s$ on this contour -
\begin{equation}
    s = a(1+i\tan\theta), \quad \theta \in (-\pi/2, \pi/2)~.
\end{equation}
\noindent Then one may show the following for $\sqrt{s}$:
\begin{equation}
    \sqrt{s} = (\sqrt{a \sec\theta}) e^{i\theta/2} =: ik \implies s = -k^2
\end{equation}
where $k$ is a complex number. Following the above representation in $\theta$ it can be shown that $k$ lies on yet another complex contour $\partial D^-$ in the lower half complex plane defined by:
\begin{equation}
    k_I^2 - k_R^2 = a, \text{ and } k_I < 0
\end{equation}
where $k_R, k_I$ are the real and imaginary parts of $k$ respectively.
\noindent Thus we have now
\begin{equation}
    \chi(t;\alpha,\gamma) = \frac{i}{\pi } \int_{\partial D^-} \frac{e^{-k^2 t} k }{-k^2 + ik\gamma +\alpha} \:dk~.
\end{equation}
Further, the above integral can be deformed to the real line using the Cauchy's integral theorem and an application of Jordan's lemma. It involves using the information that none of the poles of the complex integrand for $\alpha, \gamma>0$ lie in the lower half of the complex $k-$ plane. Finally, we have:
\begin{equation}
    \chi(t;\alpha,\gamma) = \frac{i}{\pi} \int_{-\infty}^{\infty}  \frac{e^{-k^2 t} k }{-k^2 + ik\gamma +\alpha} \:dk,
\end{equation}
which is the claim we made in Eq. \ref{chiform: laplace}.
 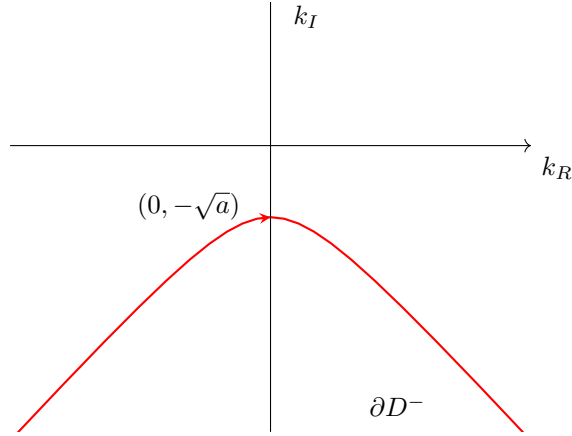
\begin{figure}[h!]
       \centering
       \begin{tikzpicture}
         \begin{axis}[hide axis,
            xmin=-4,xmax=4,
            ymin=-4,ymax=2]
            \addplot [red,thick,domain=-2.5:2.5,decorated arrows] ({sinh(x)}, -{cosh(x)});
            \addplot [black,domain=-4:4, ->] ({x},{x*0});
            \addplot [black,domain=-4:4, ->] ({x*0},{x});
         \end{axis}
         \node at (2.35,3) {$(0,-\sqrt{a})$};
         \node at (5.1, 0.345) {$\partial D^-$};
         \node at (7.2,3.5) {$k_R$};
         \node at (3.9,5.5) {$k_I$};
       \end{tikzpicture}
       \caption{Contour $\partial D^-$ (continuous red).}
\end{figure}

\section{Action of nonlinearity \texorpdfstring{$N_{\alpha}$}{} on the numerical solution}
\label{A2}
\noindent We want to estimate $N_{\alpha}(w_{ni},\tau_{ni})$ in terms of  the known quantities computed at $t_n$. We assumed the smooth mapping of the function $N_{\alpha}(w(t),t)$ to $\tilde{N}_{\alpha}(t)$. This does not hold when the first argument is replaced by a quantity different from $w(t)$. This is indeed the case in the numerical scheme since the nonlinearity acts on $w_{ni}$ which is only an approximation to the true variable $w(t_{ni})$. We estimate then to what degree this approximation affects the nonlinear action. We recall that $w_{ni} = w(\tau_{ni}) + E_{ni}$ where $E_{ni}$ is the stage error function. We have
\begin{equation}
    N_{\alpha}(w_{ni},\tau_{ni}) = N_{\alpha}(w(\tau_{ni})+ E_{ni},\tau_{ni} )~.
\end{equation}
We Taylor-expand the nonlinear function around $(w(\tau_{ni}), \tau_{ni})$ assuming it is well-behaved which yields,
\begin{equation}
\begin{aligned}
     N_{\alpha}(w(\tau_{ni})+E_{ni},\tau_{ni}) &= N_{\alpha}(w(\tau_{ni}),\tau_{ni}) + \frac{\partial N_{\alpha}}{\partial w}\Big|_{w(\tau_{ni}),\tau_{ni}} E_{ni} + \mathcal{O}(E_{ni}^2),\\
     &= \tilde{N}_{\alpha}(\tau_{ni}) + \frac{\partial N_{\alpha}}{\partial w}\Big|_{w(\tau_{ni}),\tau_{ni}} E_{ni} + \mathcal{O}(E_{ni}^2),
\end{aligned}
\end{equation}
where we have now used the smooth mapping function argument to replace the first term on the right-hand size with $\tilde{N}_{\alpha}$.\\
\indent Now, the derivative $\partial N_{\alpha}/\partial w$ is yet another function of $w$ such that we can define $G(w(t)):=\partial N_{\alpha}/\partial w$. Further, we assume a smooth mapping to exist such that $G(w(t)) = \tilde{G}(t)$. This function too must inherit the behaviour of $N_{\alpha}$ such that a similar series expansion as in Eq. \ref{seriesexp} must exist. This allows us to write:
\begin{equation}
    \frac{\partial N_{\alpha}}{\partial w}\Big|_{w(\tau_{ni}),\tau_{ni}}  = G(w(\tau_{ni})) = \tilde{G}(\tau_{ni}) = \tilde{G}(t_n) +c \sqrt{\Delta t} + ...
\end{equation}
for some constant, $c$. Thus we have,
\begin{equation}
     N_{\alpha}(w(\tau_{ni})+E_{ni},\tau_{ni})  = \tilde{N}_{\alpha}(\tau_{ni}) + \tilde{G}(t_n) E_{ni} + \mathcal{O}(E_{ni}\sqrt{\Delta t})
\end{equation}
which is the estimate claimed in Eq. \ref{eq:Nexpansion} once we identify $\tilde{G}=J_n$.

\section{Error estimate in the history function integral}
\label{A3}

\noindent We want to verify the following estimate for both the 2-stage and 4-stage schemes:
\begin{equation}{\label{Ctoshow}}
    \rho(k) \sum_{m=0}^n\Bigg(\Delta t \sum_{i=1}^s d_i(k) J_m \delta_{mi}\Bigg) \sim \mathcal{O}(\Delta t^{s/2})~.
\end{equation}
We have already established the following involving the $\{b_i\}$'s:
\[\Delta t \sum_{i=1}^s b_i J_m \delta_{mi} \sim O(\Delta t^{s/2+1})~.\]
However, we want to show the convergence behaviour of a similar expression as above  with respect to $\Delta t$ with $\{d_i(k)\}$ instead of $\{b_i\}$. Recall that $d_i$'s have been fixed already at this point via the order conditions Eqs. \ref{oc2:2f} and \ref{oc2:4f}, so we can no longer impose any further independent conditions on them. What we can do instead is only assess how the sum behaves given the $\{d_i\}$'s. Recall also that $\delta_{n1} = 0$ for all $n$. We will verify separately for $s=2$ and $s=4$ that \ref{Ctoshow} holds. We introduce a shorthand notation for the expressions 
\begin{equation}
    \sum_{i=1}^{j-1}a_{ji}c_i^{\frac{l-1}{2}} - c_j^{\frac{l+1}{2}} \phi_{(l-1)/2,j}(\tilde{\gamma}) =: [OC]_{j,l}
\end{equation}
which become order conditions when they are zero for some $j,l$.
\subsection{s=2}
\begin{equation}
    \begin{aligned}
        J_m \Delta t \sum_{i=1}^{s=2} d_i(k) \delta_{mi}&= (J_m \Delta t )d_2(k) \delta_{m2}=(J_m \Delta t)d_2(k) \sum_{l=1}^{p_{2,2}=1}C_{l-1}(t_m) \Delta t^{\frac{l+1}{2}} [OC]_{2,l} + \mathcal{O}(\Delta t^{q(m)})~.
    \end{aligned}
\end{equation}
 In the two-stage scheme, $[OC]_{2,1}=0$ via Eq. \ref{oc3:2f}. The leading behaviour of the remainder however is time-dependent. We recall that the coefficients $C_l$ for the odd index are zero for $t>0$. This means that for $m=0$, we have $q(0) = 5/2$ while for $m>0$, we have $q(m) = 3$. As a result, for large enough $n$,
\[\sum_{m=0}^n J_m \Delta t \sum_{i=1}^2 d_i(k) \delta_{mi} \sim \mathcal{O}(\Delta t^2)\]
which is a better estimate than required in the relation \ref{Ctoshow}.
\subsection{s=4}
\begin{equation}{
\label{eq:C4}
}
    \begin{aligned}
        J_m \Delta t\sum_{i=1}^4 d_i(k) \delta_{mi} &\sim J_m \Delta t \sum_{i=2}^4d_i(k) \sum_{l=1}^{p_{2,i}} C_{l-1}(t_m)  (\Delta t)^{\frac{l+1}{2}}[OC]_{i,l}
    \end{aligned}
\end{equation}
Here, the leading contribution  proportional to $C_1(t_m) (\Delta t)^{5/2}$ comes due to non-zero expressions $[OC]_{i,2}$. Once again, this contribution becomes zero for $m>0$. Note that the other expressions $[OC]_{i,l}$ in Eq. \ref{eq:C4} are zero due to the order conditions Eqs. \ref{oc3:4f} and \ref{oc5:4f}. Thus we have,
\[\sum_{m=0}^n J_m \Delta t \sum_{i=1}^4 d_i(k) \delta_{mi} \sim \mathcal{O}(\Delta t^2)\]
which verifies \ref{Ctoshow} for $s=4$.

\bibliography{rkxtd-main}{}

\begin{thebibliography}{25}
\expandafter\ifx\csname natexlab\endcsname\relax\def\natexlab#1{#1}\fi
\providecommand{\url}[1]{\texttt{#1}}
\providecommand{\href}[2]{#2}
\providecommand{\path}[1]{#1}
\providecommand{\DOIprefix}{doi:}
\providecommand{\ArXivprefix}{arXiv:}
\providecommand{\URLprefix}{URL: }
\providecommand{\Pubmedprefix}{pmid:}
\providecommand{\doi}[1]{\href{http://dx.doi.org/#1}{\path{#1}}}
\providecommand{\Pubmed}[1]{\href{pmid:#1}{\path{#1}}}
\providecommand{\bibinfo}[2]{#2}
\ifx\xfnm\relax \def\xfnm[#1]{\unskip,\space#1}\fi
\bibitem[{{M}axey and {R}iley(1983)}]{Maxey1983}
\bibinfo{author}{M.~R. {M}axey}, \bibinfo{author}{J.~J. {R}iley},
\newblock \bibinfo{title}{Equation of motion for a small rigid sphere in a
  nonuniform flow},
\newblock \bibinfo{journal}{Phys. Fluids} \bibinfo{volume}{26}
  (\bibinfo{year}{1983}) \bibinfo{pages}{883--889}.
\bibitem[{Gatignol(1983)}]{gatignol1983}
\bibinfo{author}{R.~Gatignol},
\newblock \bibinfo{title}{The {F}a\'xen formulae for a rigid particle in an
  unsteady non-uniform {S}tokes flow},
\newblock \bibinfo{journal}{J. Mec. Theor. Appl.} \bibinfo{volume}{2}
  (\bibinfo{year}{1983}) \bibinfo{pages}{241--282}.
\bibitem[{Auton et~al.(1988)Auton, Hunt, and Prud'Homme}]{auton1988}
\bibinfo{author}{T.~R. Auton}, \bibinfo{author}{J.~C.~R. Hunt},
  \bibinfo{author}{M.~Prud'Homme},
\newblock \bibinfo{title}{The force exerted on a body in inviscid unsteady
  non-uniform rotational flow},
\newblock \bibinfo{journal}{J. Fluid Mech.} \bibinfo{volume}{197}
  (\bibinfo{year}{1988}) \bibinfo{pages}{241–257}.
\bibitem[{{B}asset(1888)}]{Basset1888}
\bibinfo{author}{A.~{B}asset}, \bibinfo{title}{Treatise on Hydrodynamics},
  \bibinfo{publisher}{Deighton, Bell and Company}, \bibinfo{year}{1888}.
\bibitem[{{B}oussinesq(1885)}]{Boussinesq1885}
\bibinfo{author}{J.~{B}oussinesq},
\newblock \bibinfo{title}{Sur la résistance qu’oppose un liquide indéfini
  au repos au mouvement varié d’une sphère solide},
\newblock \bibinfo{journal}{C. R. Acad. Sci. Paris} \bibinfo{volume}{100}
  (\bibinfo{year}{1885}) \bibinfo{pages}{935–937}.
\bibitem[{Farazmand and Haller(2015)}]{Farazmand15}
\bibinfo{author}{M.~Farazmand}, \bibinfo{author}{G.~Haller},
\newblock \bibinfo{title}{The {M}axey–{R}iley equation: Existence, uniqueness
  and regularity of solutions},
\newblock \bibinfo{journal}{Nonlinear Anal. Real World Appl.}
  \bibinfo{volume}{22} (\bibinfo{year}{2015}) \bibinfo{pages}{98--106}.
\bibitem[{Haller(2019)}]{haller2019}
\bibinfo{author}{G.~Haller},
\newblock \bibinfo{title}{Solving the inertial particle equation with memory},
\newblock \bibinfo{journal}{J. Fluid Mech.} \bibinfo{volume}{874}
  (\bibinfo{year}{2019}) \bibinfo{pages}{1–4}.
\bibitem[{Brush et~al.(1964)Brush, Ho, and Yen}]{brush1964}
\bibinfo{author}{L.~M. Brush}, \bibinfo{author}{H.-W. Ho},
  \bibinfo{author}{B.-C. Yen},
\newblock \bibinfo{title}{Accelerated motion of a sphere in a viscous fluid},
\newblock \bibinfo{journal}{J. Hydraul. Eng.} \bibinfo{volume}{90}
  (\bibinfo{year}{1964}) \bibinfo{pages}{149--160}.
\bibitem[{{van Hinsberg} et~al.(2011){van Hinsberg}, {ten Thije Boonkkamp}, and
  Clercx}]{vanhinsberg2011}
\bibinfo{author}{M.~{van Hinsberg}}, \bibinfo{author}{J.~{ten Thije
  Boonkkamp}}, \bibinfo{author}{H.~Clercx},
\newblock \bibinfo{title}{An efficient, second order method for the
  approximation of the {B}asset history force},
\newblock \bibinfo{journal}{J. Comput. Phys.} \bibinfo{volume}{230}
  (\bibinfo{year}{2011}) \bibinfo{pages}{1465--1478}.
\bibitem[{Daitche(2013)}]{daitche2013}
\bibinfo{author}{A.~Daitche},
\newblock \bibinfo{title}{Advection of inertial particles in the presence of
  the history force: Higher order numerical schemes},
\newblock \bibinfo{journal}{J. Comput. Phys.} \bibinfo{volume}{254}
  (\bibinfo{year}{2013}) \bibinfo{pages}{93--106}.
\bibitem[{Dorgan and Loth(2007)}]{dorganloth2007}
\bibinfo{author}{A.~Dorgan}, \bibinfo{author}{E.~Loth},
\newblock \bibinfo{title}{Efficient calculation of the history force at finite
  {R}eynolds numbers},
\newblock \bibinfo{journal}{Int. J. Multiphase Flow} \bibinfo{volume}{33}
  (\bibinfo{year}{2007}) \bibinfo{pages}{833--848}.
\bibitem[{Bombardelli et~al.(2008)Bombardelli, Gonz{\'{a}}lez, and
  Ni{\~{n}}o}]{bombardelli2008}
\bibinfo{author}{F.~Bombardelli}, \bibinfo{author}{A.~Gonz{\'{a}}lez},
  \bibinfo{author}{Y.~Ni{\~{n}}o},
\newblock \bibinfo{title}{Computation of the particle {B}asset force with a
  fractional-derivative approach},
\newblock \bibinfo{journal}{J. Hydraul. Eng.} \bibinfo{volume}{134}
  (\bibinfo{year}{2008}) \bibinfo{pages}{1513--1520}.
\bibitem[{Parmar et~al.(2018)Parmar, Annamalai, Balachandar, and
  Prosperetti}]{parmar2018}
\bibinfo{author}{M.~Parmar}, \bibinfo{author}{S.~Annamalai},
  \bibinfo{author}{S.~Balachandar}, \bibinfo{author}{A.~Prosperetti},
\newblock \bibinfo{title}{Differential formulation of the viscous history force
  on a particle for efficient and accurate computation},
\newblock \bibinfo{journal}{J. Fluid Mech.} \bibinfo{volume}{844}
  (\bibinfo{year}{2018}) \bibinfo{pages}{970–993}.
\bibitem[{Siegle et~al.(2011)Siegle, Goychuk, and
  Hänggi}]{markovianEmbedding_Siegle2011}
\bibinfo{author}{P.~Siegle}, \bibinfo{author}{I.~Goychuk},
  \bibinfo{author}{P.~Hänggi},
\newblock \bibinfo{title}{{M}arkovian embedding of fractional superdiffusion},
\newblock \bibinfo{journal}{Europhys. Lett.} \bibinfo{volume}{93}
  (\bibinfo{year}{2011}) \bibinfo{pages}{20002}.
\bibitem[{McCloskey and Paternostro(2014)}]{markovianEmbedding1}
\bibinfo{author}{R.~McCloskey}, \bibinfo{author}{M.~Paternostro},
\newblock \bibinfo{title}{Non-{M}arkovianity and system-environment
  correlations in a microscopic collision model},
\newblock \bibinfo{journal}{Phys. Rev. A} \bibinfo{volume}{89}
  (\bibinfo{year}{2014}) \bibinfo{pages}{052120}.
\bibitem[{Kretschmer et~al.(2016)Kretschmer, Luoma, and
  Strunz}]{markovianEmbedding2}
\bibinfo{author}{S.~Kretschmer}, \bibinfo{author}{K.~Luoma},
  \bibinfo{author}{W.~T. Strunz},
\newblock \bibinfo{title}{Collision model for non-{M}arkovian quantum
  dynamics},
\newblock \bibinfo{journal}{Phys. Rev. A} \bibinfo{volume}{94}
  (\bibinfo{year}{2016}) \bibinfo{pages}{012106}.
\bibitem[{Campbell et~al.(2018)Campbell, Ciccarello, Palma, and
  Vacchini}]{markovianEmbedding_campbell}
\bibinfo{author}{S.~Campbell}, \bibinfo{author}{F.~Ciccarello},
  \bibinfo{author}{G.~M. Palma}, \bibinfo{author}{B.~Vacchini},
\newblock \bibinfo{title}{System-environment correlations and {M}arkovian
  embedding of quantum non-{M}arkovian dynamics},
\newblock \bibinfo{journal}{Phys. Rev. A} \bibinfo{volume}{98}
  (\bibinfo{year}{2018}) \bibinfo{pages}{012142}.
\bibitem[{Prasath et~al.(2019)Prasath, Vasan, and Govindarajan}]{sgp2019}
\bibinfo{author}{S.~G. Prasath}, \bibinfo{author}{V.~Vasan},
  \bibinfo{author}{R.~Govindarajan},
\newblock \bibinfo{title}{Accurate solution method for the {M}axey-{R}iley
  equation, and the effects of {B}asset history},
\newblock \bibinfo{journal}{J. Fluid Mech.} \bibinfo{volume}{868}
  (\bibinfo{year}{2019}) \bibinfo{pages}{428--460}.
\bibitem[{Garrappa and Popolizio(2011)}]{Garrappa2011}
\bibinfo{author}{R.~Garrappa}, \bibinfo{author}{M.~Popolizio},
\newblock \bibinfo{title}{Generalized exponential time differencing methods for
  fractional order problems},
\newblock \bibinfo{journal}{Comput. Math. Appl.} \bibinfo{volume}{62}
  (\bibinfo{year}{2011}) \bibinfo{pages}{876--890}.
\bibitem[{Garrappa(2012)}]{garrappa2012stability}
\bibinfo{author}{R.~Garrappa},
\newblock \bibinfo{title}{Stability-preserving high-order methods for multiterm
  fractional differential equations},
\newblock \bibinfo{journal}{Int. J. Bifurcation Chaos} \bibinfo{volume}{22}
  (\bibinfo{year}{2012}) \bibinfo{pages}{1250073}.
\bibitem[{Cox and Matthews(2002)}]{CM2002}
\bibinfo{author}{S.~Cox}, \bibinfo{author}{P.~Matthews},
\newblock \bibinfo{title}{Exponential time differencing for stiff systems},
\newblock \bibinfo{journal}{J. Comput. Phys.} \bibinfo{volume}{2}
  (\bibinfo{year}{2002}) \bibinfo{pages}{430--455}.
\bibitem[{Hochbruck and Ostermann(2005)}]{HO2005}
\bibinfo{author}{M.~Hochbruck}, \bibinfo{author}{A.~Ostermann},
\newblock \bibinfo{title}{Explicit exponential {R}unge-{K}utta methods for
  semilinear parabolic problems},
\newblock \bibinfo{journal}{SIAM J. Numer. Anal.} \bibinfo{volume}{43}
  (\bibinfo{year}{2005}) \bibinfo{pages}{1069--1090}.
\bibitem[{Curtain and Zwart(2012)}]{curtain2012introduction}
\bibinfo{author}{R.~F. Curtain}, \bibinfo{author}{H.~Zwart}, \bibinfo{title}{An
  introduction to infinite-dimensional linear systems theory},
  volume~\bibinfo{volume}{21}, \bibinfo{publisher}{Springer Science \& Business
  Media}, \bibinfo{year}{2012}.
\bibitem[{Langlois et~al.(2015)Langlois, Farazmand, and Haller}]{Langlois15}
\bibinfo{author}{G.~Langlois}, \bibinfo{author}{M.~Farazmand},
  \bibinfo{author}{G.~Haller},
\newblock \bibinfo{title}{Asymptotic dynamics of inertial particles with
  memory},
\newblock \bibinfo{journal}{J. Nonlinear Sci.} \bibinfo{volume}{25}
  (\bibinfo{year}{2015}) \bibinfo{pages}{1225–1255}.
\bibitem[{Lovalenti and Brady(1993)}]{LO1993}
\bibinfo{author}{P.~M. Lovalenti}, \bibinfo{author}{J.~F. Brady},
\newblock \bibinfo{title}{The hydrodynamic force on a rigid particle undergoing
  arbitrary time-dependent motion at small {R}eynolds number},
\newblock \bibinfo{journal}{J. Fluid Mech.} \bibinfo{volume}{256}
  (\bibinfo{year}{1993}) \bibinfo{pages}{561–605}.

\end{thebibliography}
\bibliographystyle{elsarticle-num-names} 
\end{document}